\documentclass{aa}  

\usepackage{graphicx}
\usepackage{txfonts}
\usepackage{lipsum}
\usepackage{subcaption}         
\usepackage{lscape}             
\usepackage{placeins}
\usepackage{hyperref}

\begin{document}

   \title{Long-lived coronal loops in solar active regions\thanks{Movies associated to Figs. 2 and 3 are available at https://www.aanda.org}}

   \author{%
   N.~Vasantharaju\inst{1}
   \and
   H.~Peter\inst{1,2}
   \and
   L.P.~Chitta\inst{1}
   \and
   S.~Mandal\inst{1}
   }

   \institute{%
    Max Planck Institute for Solar System Research, 
    Justus-von-Liebig-Weg 3, 37077 Göttingen, Germany;
    \\ \email{naganna@mps.mpg.de}
    \and
    Institut f\"ur Sonnenphysik (KIS), Georges-K\"ohler-Allee 401A, 79110 Freiburg, Germany
    }

   \date{Received 4 July 2025 / Accepted 14 September 2025}

  \abstract
 { Coronal loops are plasma structures in the solar atmosphere with temperatures reaching millions of Kelvin, shaped and sustained by the magnetic field. However, their morphology and fundamental nature remain subjects of debate. By studying their cross-sectional properties and how they change along the loop and in time, we can understand their magnetic structure and heating mechanisms. In this study, we investigated the cross-sectional intensity profiles, both spatially and temporally, of two unique coronal loops, observed in the periphery of two distinct active regions by the Extreme Ultraviolet Imager (EUI/HRI$_{\rm EUV}$) on board the Solar Orbiter spacecraft. The main results of this study are 1. The lifetimes of these two loops (loop1 > 120 min \& loop2 > 50 min) are longer than the typical timescales of radiative cooling and thermal conduction. 2. Their widths determined by the FWHM of the single Gaussian fit to the cross-axis intensity profiles are greater than 6-7 pixels of EUI/HRI$_{\rm EUV}$, indicating that the loop cross-section is uniformly filled on well-resolvable scales. 3. These loops exhibited an almost constant width, both spatially and temporally (width for loop1 is 2.1 $\pm$ 0.4 Mm and for loop2 is 1.3 $\pm$ 0.2 Mm), indicating that they are stable non-expanding structures. 4. We present observational evidence that the one of the loops (loop2) is not braided, which strongly suggests that the non-expanding nature of this multi-stranded loop along its length cannot be attributed to the twist of the magnetic field lines. In conclusion, we find that these coronal loops are long, stable, multi-stranded, non-expanding structures with a uniform cross-section that persist in the corona for an unusually extended duration. This not only challenges our current understanding of the structure of the coronal magnetic field but also raises critical questions about the mechanisms responsible for the remarkable stability of these loops.}


   \maketitle

\section{Introduction}
Coronal loops are bright, arch-like structures in the solar corona composed of hot plasma ($10^6 - 10^7$ K), shaped by the  magnetic field \citep{Fabio2014}. The cross-sectional intensity profiles of coronal loops provide valuable information about the distribution of plasma and the configuration of the magnetic field within these structures. Since direct measurements of coronal magnetic fields are very difficult, the analysis of the cross-sectional properties of loops, e.g. their strandedness and state of braiding (\cite{Chitta2022}), is rather important to understand the structure of the coronal magnetic field and the spatial distribution of coronal heating.

Previous observations (e.g., \citealp{Klimchuk2000a, Watko2000}) have shown that most coronal loops maintain an almost constant cross-section along their length or exhibit only minimal expansion with height. However, magnetic field extrapolations from the photospheric surface to the corona reveal significant expansion of the field \citep{Fuentes2006}. Given that coronal loops trace magnetic field lines, it would be expected that they should also expand with height, which is not supported by the observations. \citet{Peter2012} attributed this discrepancy to the varying contributions of the loop plasma to the emission at different temperatures. Several other studies have proposed different explanations to address this discrepancy, as discussed in the introduction of \citet{Peter2012}. That a combination of magnetic structure and viewing angle might give the illusion of a loop of constant cross section while the field is actually expanding has been studied by  \cite{2013ApJ...775..120M} and \cite{2022ApJ...927....1M}. Another possible explanation, relevant to the present study, is that a significant twist in the magnetic flux tube suppresses the apparent expansion when viewed from the side \citep{Klimchuk2000b}. This effect can manifest as either the twisting of an individual loop or the 'braiding' of a bundle of unresolved, thin loop strands. Despite these proposed solutions, the observed near-constant thickness of coronal loops still remains an open question as these solutions cannot explain all the observed properties of the loops.

Another important aspect of a coronal loop is its observed lifetime. Understanding the mechanisms responsible for loop longevity also helps in constraining coronal-heating theories. Generally, the lifetime of a coronal loop is governed by the coronal cooling time, which depends on thermal conduction and radiative losses. Under typical coronal conditions, this cooling time is estimated to be around 20–30 minutes \citep{Rosner1978,Peter2012}. If a loop's lifetime closely matches its cooling time, it likely has a monolithic structure, allowing it to cool as a homogeneous unit. In contrast, if a loop persists significantly longer than this cooling time, there are (at least) two possibilities. The loop could be heated steadily (or at a high enough frequency), or the loop could be composed of multiple unresolved fine strands, with individual strands undergoing episodic heating events at different times. This scenario allows the overall structure to appear stable over extended durations \citep{Amy2003}. \citet{Klimchuk2010} supported this notion using 1D hydrodynamic modeling, demonstrating that multi-stranded loop models align better with observed loop properties than monolithic models. Additionally, individual strands within a loop experience independent heating and cooling cycles, consistent with observations of multi-thermal plasma distributions in loops with extended lifetimes \citep{Warren2008}.

Many past observations have reported loops with lifetimes longer than the characteristic cooling time (e.g.~\citealp{Fuentes2007,Warren2002,Warren2003}). However, many of these observations were made with limited resolution and using different instruments and/or filters, often with considerable time delays between loop appearances. For example, \citet{Amy2005} showed that hot monolithic loops visible in X-ray images from the Yohkoh instrument were later resolved as cooler stranded structures in EUV images from the TRACE instrument, with a time delay of 1 to 3 hours. Consequently, establishing a direct correlation between the loops detected by Yohkoh and those detected by TRACE is challenging. Recently, \citet{Li2020} reported a loop with a lifetime of about 90 minutes using observations from AIA onboard the SDO. However, a detailed study of the cross-sectional properties of loops over their extended lifetimes remains lacking.

In this new study we investigate the spatial and temporal variation of cross-sectional widths of two unique loops, observed with EUI on board Solar Orbiter, over their extended lifetimes. By investigating the cross-sectional evolution of loops, this study aims to shed light on their morphology, internal structure, stability, and underlying heating mechanisms. 

The paper is organized as follows. Sect.~\ref{obs} presents the observations of loops obtained with EUI/HRI$_{\rm EUV}$. Sect.~\ref{res} describes the loop morphology and internal structure. Sect.~\ref{dis} discusses possible heating scenarios that may account for the extended lifetimes of the loops, as well as the non-braided nature of their internal structure. Finally, Sect.~\ref{summ} summarizes our conclusions.

\section{Observations}
\label{obs} 
The Extreme Ultraviolet Imager (EUI, \citealp{Rochus2020}) onboard Solar Orbiter \citep{2020A&A...642A...1M} consists of three telescopes: one Full Sun Imager (FSI) and two High Resolution Imagers (HRIs). The FSI provides global-scale images of the Sun in the 174~\AA\ (FSI$_{174}$) and 304~\AA\ (FSI$_{304}$) EUV passbands. The HRIs deliver high-resolution images of the solar atmosphere: one in the 174~\AA\ EUV passband (HRI$_{\rm EUV}$) and another in the Lyman-$\alpha$ line at 1215~\AA\ (HRI$_{\rm Lya}$). In this study, we use sets of calibrated Level-2 HRI$_{\rm EUV}$ data\footnote{\url{https://www.sidc.be/EUI/data/releases/202301_release_6.0}} which contain two unique coronal loops observed on 21 August 2021 and 3 April 2022. The HRI$_{\rm EUV}$ images the solar plasma at temperatures of approximately 1~MK, with a plate scale of 0.492{\arcsec} per pixel. During the observations used in this study, the Solar Orbiter was at distances of 0.653 AU and 0.363 AU from the Sun on 21 August 2021 and 3 April 2022 respectively, making each pixel of HRI$_{\rm EUV}$ equivalent to 230 km and 128 km on the Sun. Similarly, FSI$_{174}$ with a plate scale of 4.44{\arcsec} per pixel has an equivalent pixel size of 2.08 Mm and 1.16 Mm on the Sun on these dates. No complementary EUV observations from Earth-orbiting spacecraft were available for this study, as the observed regions were on the far side of the Sun as seen from Earth. This limitation is due to the large angular separations between Solar Orbiter and the Sun-Earth line, which are approximately $-81^{\circ}$ on 21 August 2021 and $110^{\circ}$ on 3 April 2022. 
 
\begin{figure*}
        \centering
       \includegraphics[trim=30 200 30 200,clip,width=\hsize]{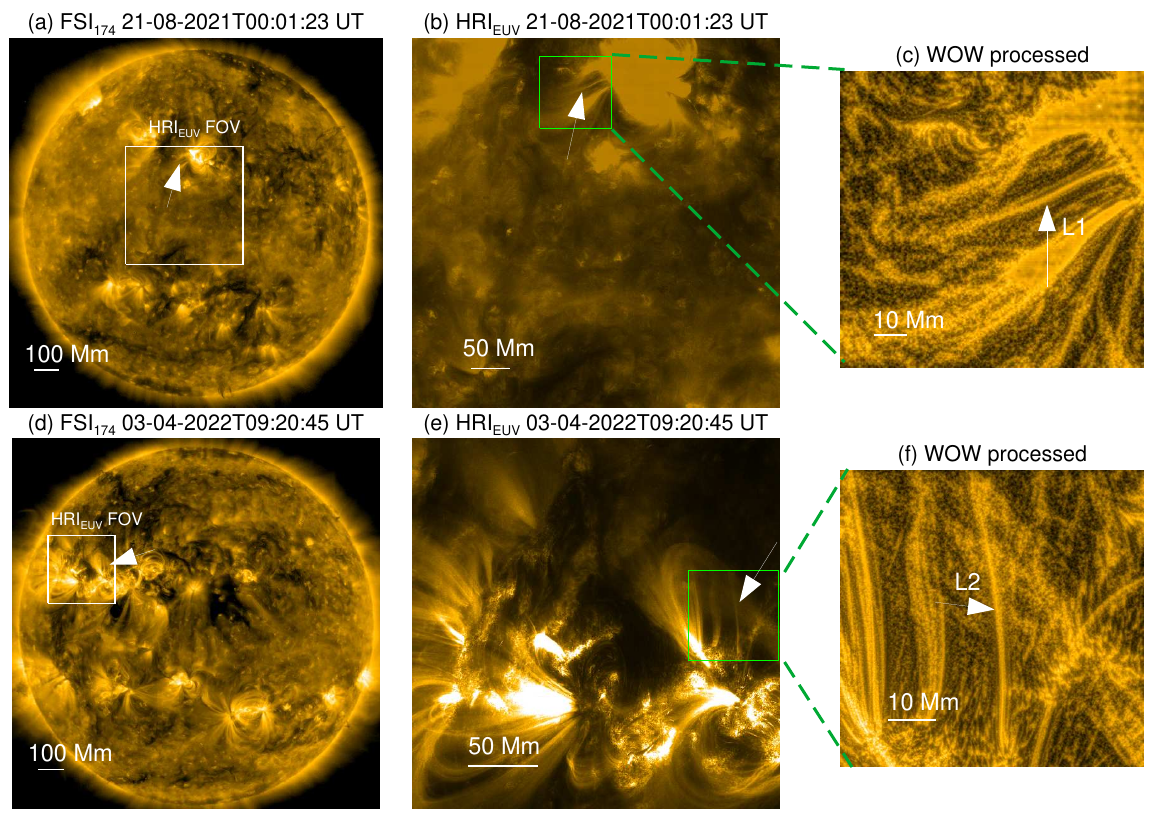}
        \caption{Overview of Loops L1 and L2. (a) Full-disk image of FSI$_{174}$, highlighting the location of L1 (indicated by a white arrow) in NOAA AR 12859. The region enclosed by the white square marks the field of view of HRI$_{\rm EUV}$ (b) Snapshot of HRI$_{\rm EUV}$, with a more zoomed-in view of L1. The region enclosed by the green square is WOW-processed and shown in (c). (d)–(f) Same as (a)–(c), but for L2.}
        \label{Fig1}%
    \end{figure*}

We investigated both loops in depth, and they are located at the periphery of active regions. Figure~\ref{Fig1} provides an overview of the two loops, named L1 and L2, indicated by white arrows. We have more than two and a half hours of HRI$_{\rm EUV}$ observations of loop L1 at a cadence of 5~s for the first hour and, for the remaining period, at a cadence of 15~s. For loop L2, we have more than an hour of continuous HRI$_{\rm EUV}$ observations at a cadence of 10~s. The jitter in the HRI$_{\rm EUV}$ images is removed using a routine \texttt{fg\_rigidalign.pro}\footnote{The routine performs cross-correlation based image alignment and is available in Solarsoft (https://www.lmsal.com/solarsoft/).}. We do not know how long these loops existed prior to the start of the HRI$_{\rm EUV}$ observations. However, the available continuous HRI$_{\rm EUV}$ images provide sufficient evidence that the lifetime of loop L1 exceeds two hours, while that of L2 exceeds 50 minutes. Nevertheless, we limited our analysis to two hours for L1 and 47 minutes for L2 (see Sect.~\ref{res1}), as the emergence of nearby bright structures at later times could potentially affect our measurements.  

Loop L1 remains well defined for the first two hours of the observations. After this time, a bright structure begins to emerge from its southern side and progressively merges with L1. The complete merging of this bright structure with L1 takes place after about three hours of observations. Loop L2 maintains its coherence for approximately 47 minutes (09:19:15–10:06:15 UT), after which its intensity decreases steadily. By about 10:12 UT, the loop emission has diffused into the surrounding background, making it difficult to identify L2 as an isolated structure.  

The images of FSI$_{174}$ in Fig.~\ref{Fig1}(a) \& (d) show the location of L1 and L2 on the Sun and their association with the respective ARs (L1 is located on the western side of NOAA AR~12859 and L2 on the northern side of an unnumbered AR). The snapshots of HRI$_{\rm EUV}$ in Fig.~\ref{Fig1}(b) \& (e) provide the high-resolution views of L1 and L2. The small regions enclosing L1 and L2, marked by green squares, were processed for contrast enhancement using Wavelet-Optimized Whitening (WOW, \citealp{Auchere2023}). The WOW-processed images are shown in Fig.~\ref{Fig1}(c) \& (f). From the HRI$_{\rm EUV}$ observations, the loops appear long, isolated, and filled with uniform intensity along their length until the emission becomes more diffused towards the end of the loop, away from the active region. It should be noted that WOW-processed images are only used for visualization purposes and not in the loop width analysis (Sect.~\ref{res1}). Furthermore, all EUV images presented in this paper are displayed using the original HRI$_{\rm EUV}$ data values in linear (DN) scaling. No logarithmic or other nonlinear intensity transformations were applied, either in the figures or in the computation of the intensity profiles. Long-lived loops with smooth, uniform morphology such as L1 and L2 are relatively rare among coronal loops.

\section{Results and Analysis}
\label{res}

In the following, we present the analysis of two particular loops which appear stable and monolithic: they persist longer than a typical coronal cooling time and show no clear substructure, such as strands, during their evolution. This introduces a selection bias, as many loops evolve more rapidly and exhibit dynamic internal structuring. Interesting and puzzling examples of such evolving, structured loops have been discussed recently by \cite{Sudip2024,2025A&A...697A.233M}. Here, we focus on stable, monolithic loops and their properties. This motivates our selection. We discuss their morphology in Sect.~\ref{res1}, and their (lack of) internal structure in Sect.~\ref{res2}.

\subsection{Loops morphology}
\label{res1}
To investigate the structure and morphology of the loops we first turn to the cross-sectional variation of the two loops (Sects.\ \ref{l1} and \ref{l2}). We then study how the width of the two loops are distributed in space and time (Sect.\ \ref{loop.width}).

\subsubsection{Cross section variation of loop L1}
\label{l1}
\begin{figure*}
    \centering
    \includegraphics[width=0.98\textwidth]{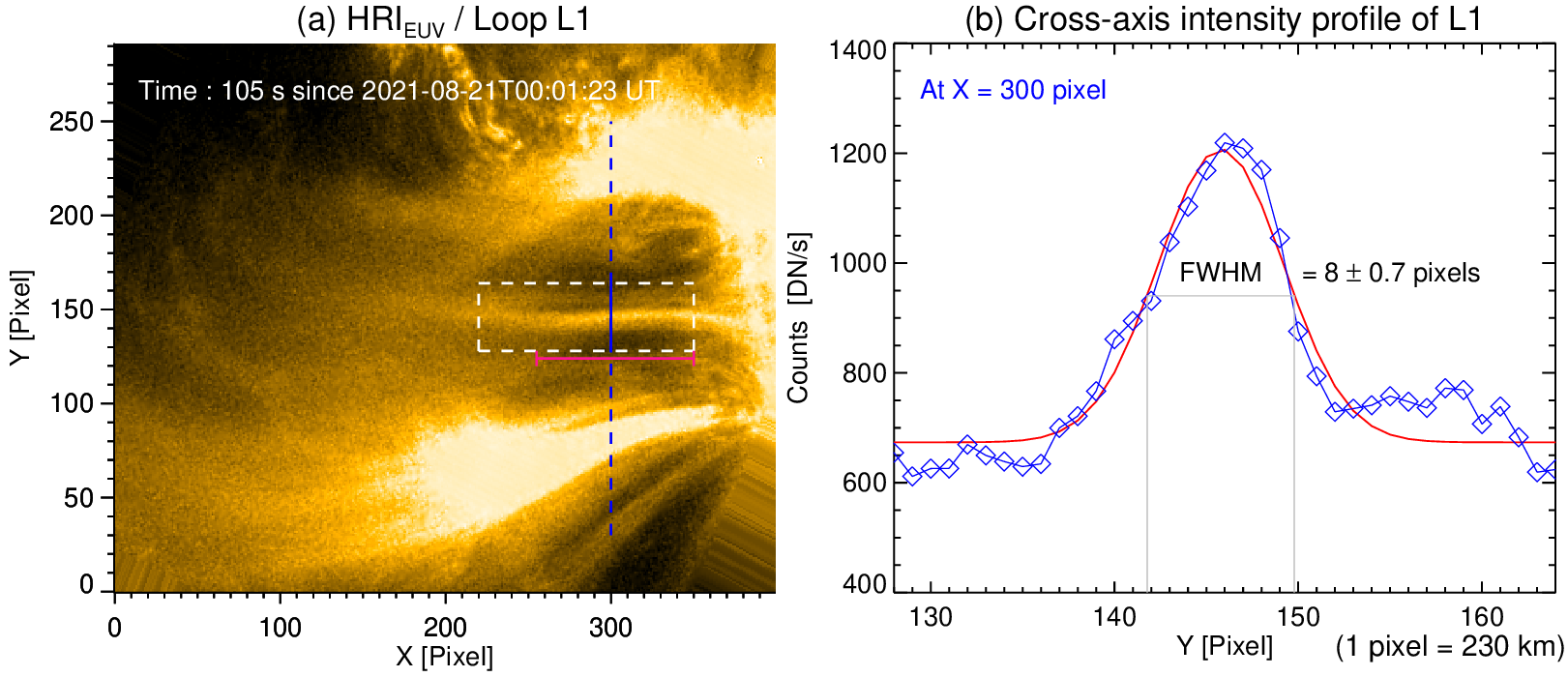}  
    \vspace{0.5cm} 
    \includegraphics[width=0.496\textwidth]{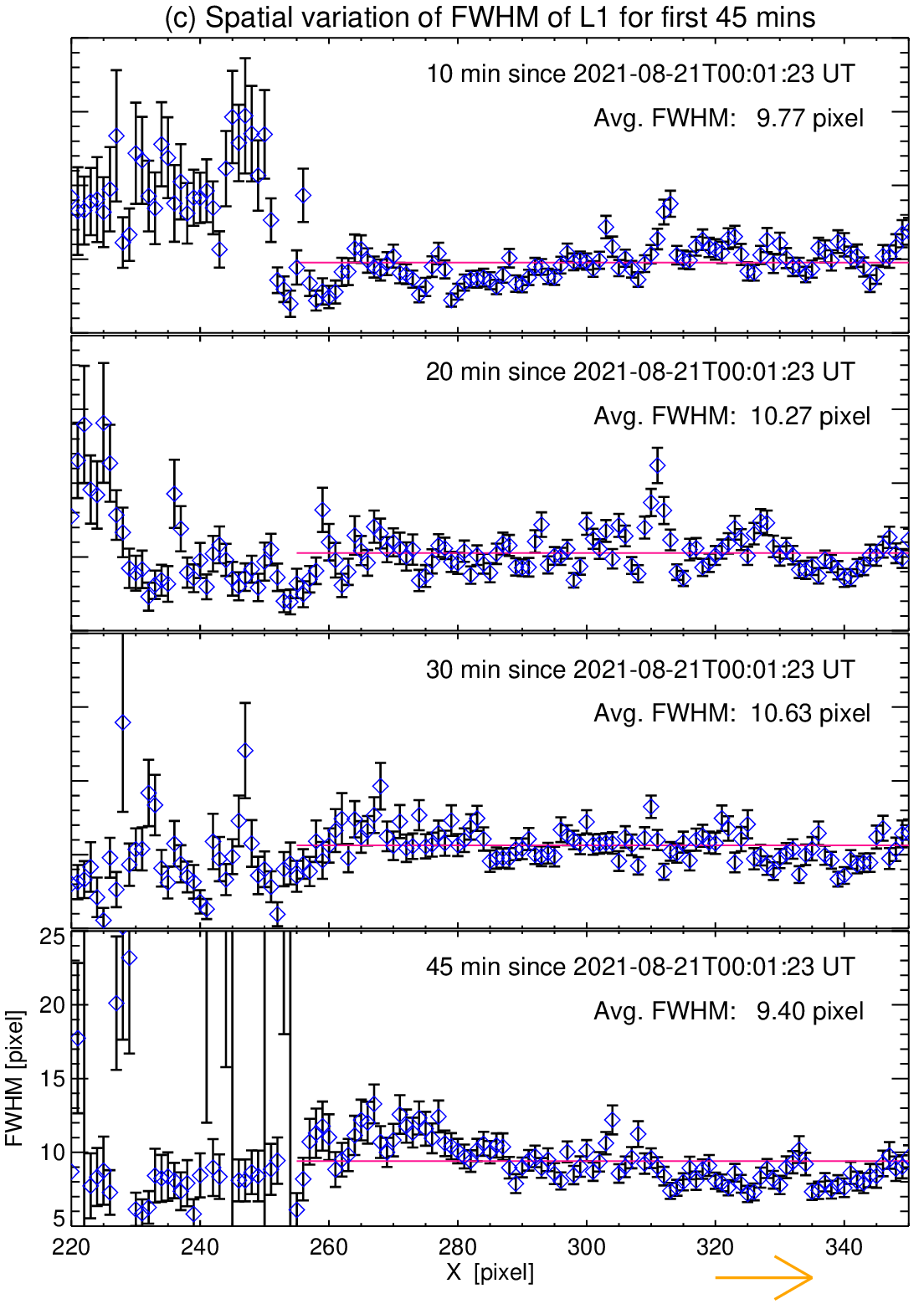}
    \hfill
    \includegraphics[width=0.496\textwidth]{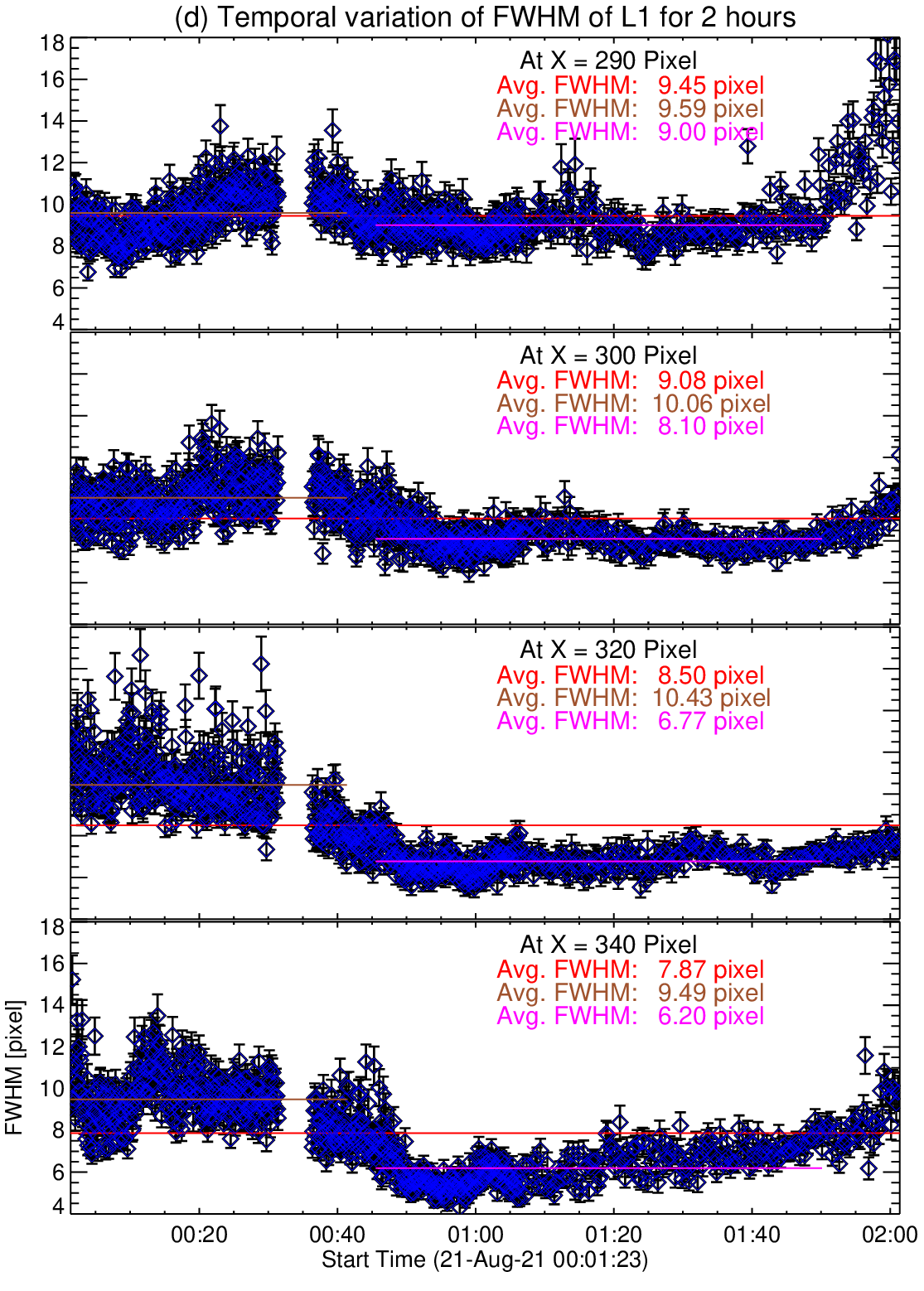} 
    \caption{(a) A sample rotated HRI$_{\rm EUV}$ snapshot with L1 aligned parallel to X-axis. The vertical blue dashed line marks the position at X=300 pixel. (b) The cross-sectional intensity profile along the vertical solid blue line in (a) is plotted as a blue curve, with a Gaussian fit overlaid in red. The FWHM of the fit is considered as width of L1 (c). Spatial variations of cross-sectional widths of Loop L1, from positions X = 220 to X = 355 pixels, are plotted at 10-minute intervals for the first 45 minutes of L1 observations. The average width is computed for the better part of L1, indicated by pink horizontal line in (a). The orange arrow at the bottom points toward the footpoint of L1. (d) Temporal variations of widths of L1 at different positions are plotted over the total two-hour period. Due to lateral disappearance of strands in L1 after 45 minutes, separate average widths for before and after this time, including the averages for whole period are noted in all panels. A short data gap occurred between 00:30 and 00:35, during which no observations were recorded.}
    \label{Fig2}
    \end{figure*}

We measured the width of the loop L1 spatially along its length over 30 Mm, and investigated its variations in time for two hours. The procedure for measuring the width of L1 is simple and can be illustrated using a sample snapshot from HRI$_{\rm EUV}$ observations of L1 (Fig.~\ref{Fig2}(a)). In the snapshot, we made L1 lie almost parallel to one axis of the image by rotating the original image, here by  about 25 degrees clockwise. In the given example, a vertical blue line is drawn perpendicular to L1 at the position X = 300 pixel and along which the generated cross-sectional intensity (CSI) profile (solid blue curve) is plotted in Fig.~\ref{Fig2}(b). The breadth (36 pixels) of the dashed white rectangle in Fig.~\ref{Fig2}(a) marks the range over which the cross-axis intensity profile of L1 is plotted. We found that most of the CSI profiles are smooth, singly peaked, and do not show clear signatures of substructure along the loop or in time. In some instances, the profiles display minor deviations from a purely Gaussian shape, such as slight asymmetries or, more rarely, multiple peaks. However, these cases are relatively uncommon, and the majority of profiles tend to be simple. For consistency, we therefore fit a single Gaussian function ($ y(x) = A \exp\left(-\frac{(x - \mu)^2}{\sigma^2}\right) + C $) to all CSI profiles $y$, as a function of distance $x$, where the parameters $A$, $\mu$, $\sigma$, $C$ are the amplitude, mean, half-width at $1/e$ of the peak of the Gaussian profile, and constant background level, respectively. The observed loops are embedded in a significant background emission, which typically amounts to 50–70 \% of the peak loop intensity, and in some cases is asymmetric across the loop cross-section (see Fig.~\ref{FigA5}, where the right-flank background is ~30\% stronger than the left). Despite these variations, the background emission is relatively stable in time over the duration of our observations, consistent with the long-lived nature of the loops themselves. When fitting Gaussian functions to the CSI profiles, the constant background term C captures the local background level at each time step. In this way, the measured full width at half maximum (FWHM) corresponds to the loop width, while the fit simultaneously accounts for the underlying emission. Appendix.~\ref{app2} illustrates the temporal evolution of the background in more detail, showing that although some fluctuations are present, no systematic variations in loop width measurements are detected.

The FWHM of the Gaussian fit is calculated as $\text{FWHM} = 2 \sigma \sqrt{ \ln 2}$ and is considered the width of L1. In Fig.~\ref{Fig2}b, the Gaussian fit in the solid red curve is over-plotted on the intensity profile and the corresponding FWHM is also shown. The fits to the intensity profiles and the associated uncertainties to the fit parameters are obtained using the routine `MPFITFUN.pro'\footnote{The routine performs the fit with user defined Gaussian function and available in Solarsoft}. In this routine, we have provided the observed intensity profile along with its error ($[{\rm{photon~noise}}^{2} + {\rm{readout~noise}}^{2}]^{1/2}$, \citealp[see e.g. Appendix A of][]{2023A&A...678A.188G}) as input and as a result the fit parameters along with their uncertainties are returned. The detector readout noise is taken as a constant value of 2~DN, while the photon noise depends on the exposure time, observed intensity (in DN~s$^{-1}$), and a constant gain factor of 6.85~DN~photon$^{-1}$. For L1 observations, which have an exposure time of 2.8~s, the estimated uncertainty in intensity ranges from 37~DN~s$^{-1}$ at the minimum observed intensity (584~DN~s$^{-1}$) to 54~DN~s$^{-1}$ at the maximum (1208~DN~s$^{-1}$). Similarly, for L2 observations with an exposure time of 1.4~s, the uncertainty ranges from 69~DN~s$^{-1}$ (at 980~DN~s$^{-1}$) to 120~DN~s$^{-1}$ (at 2945~DN~s$^{-1}$). 

The CSI profiles are generated along loop L1 from X $=$ 220 to 350 pixels ($\approx$ 30 Mm; dashed white rectangle in Fig.~\ref{Fig2}a). Although L1 extends further rightward, the segment at X $<$ 220 pixels shows insufficient contrast with the background, preventing reliable Gaussian fits. In addition, the emission between X $=$ 220–255 pixels is quite diffuse (Fig.~\ref{Fig2}a), leading to large scatter in the width measurements (Fig.~\ref{Fig2}c). To ensure robust measurements, we restrict the analysis to the clearer and more confined segment between X $=$ 255–350 pixels (95 pixels $=$ 21.85 Mm), marked by the pink line in Fig.~\ref{Fig2}a. The spatial variations of the width of L1, computed at 10-minute intervals during the first 45 minutes, are shown in four different panels of Fig.~\ref{Fig2}(c). Excluding the diffuse portion (X $<$ 255 pixels) raises the possibility that part of L1 may be expanding and losing contrast with the background. Without magnetic connectivity information from photospheric magnetograms, this cannot be confirmed. Nevertheless, in the analyzed 21.9 Mm segment, the loop width remains constant both spatially and temporally. Whether the excluded portion reflects genuine expansion or reduced visibility remains an open question. As evident in Fig.~\ref{Fig2}(c), the width of L1 tends to remain almost constant at all times for the first 45 minutes and at all positions along its length of 21.9 Mm. This result implies that the loop L1 has a remarkably stable non-expanding structure with a constant cross-sectional diameter of 10 $\pm$ 0.7 pixels (2.3 Mm) and a length of about 22 Mm that persists in the solar corona for a period of at least 45 minutes.

We further investigated the width variations of loop L1 for a total duration of two hours along the clearer segment of its length ($\approx$ 21.9 Mm). The L1 width at every 5-pixel (1.15 Mm) distance along its length is measured for two hours. Four sample plots of the temporal variations of the width at different pixel positions (X $=$ 290, 300, 320, and 340) are shown in Fig.~\ref{Fig2}(d). We found that the width of L1, particularly in the segment X $=$ 290–340 pixels, decreases by a few pixels after 45 minutes of observations. This could possibly be due to the disappearance of a few strands on one side of the loop (see attached movie). By “strand disappearance” we refer to strands that gradually fade until their intensity becomes comparable to the local background. In the segment X $=$ 290–340 pixels, a few strands on the southern side of L1 diffuse into the background. This process does not lead to a noticeable decrease in the peak intensity of the CSI profiles, but it does result in a narrowing of the loop width by $\approx$ 3 pixels after about 45 minutes. As mentioned earlier, the width of L1 remains constant, with an average of 10 pixels, during the first 45 minutes. However, after this time, the width decreases from X $=$ 290 to 340 by $\approx$ 3 pixels ($<$ 0.69 Mm). In L1, at higher positions (X $<$ 290 pixels), the emission becomes slightly diffuse after 45 minutes of observation, which may indicate loop expansion and reduced contrast with the background. Since all images are displayed and analyzed in their original linear DN scaling, this behavior cannot be attributed to nonlinear intensity transformations. Nonetheless, whether the diffuse emission corresponds to genuine expansion or simply loss of contrast with the background remains uncertain. Although L1 appears to be slightly compressed towards the footpoint, the width of L1 tends to be almost constant at the respective positions along its length for two hours of observations (Fig.~\ref{Fig2}d). The average width of L1 is always greater than 6 pixels throughout the duration of 2 hours, implying that the L1 cross section is uniformly filled on well-resolvable scales of HRI$_{\rm EUV}$.

It is important to note that we see tiny and narrow structures in other parts of the image that have a width of only about two pixels. Hence the fact that we see a relatively smooth profile with typically 8 pixels width cannot attributed to a poor spatial resolution of EUI. We therefore conclude that the observed smoothness is a genuine solar feature and not an instrumental artifact. We emphasize, however, that smooth CSI profiles should not be interpreted as evidence for circular cross-sections, since we lack stereoscopic constraints on the true loop geometry. Smooth profiles may also arise if the loops are slab-like or veil-like structures observed from a single vantage point (e.g., \citealp{2022ApJ...927....1M}). Our point here is that the profiles are mostly free of discernible substructure and show no obvious signatures of expansion.

\subsubsection{Cross section variation of loop L2}
\label{l2}
\begin{figure*}
    \centering
    \includegraphics[width=0.98\textwidth]{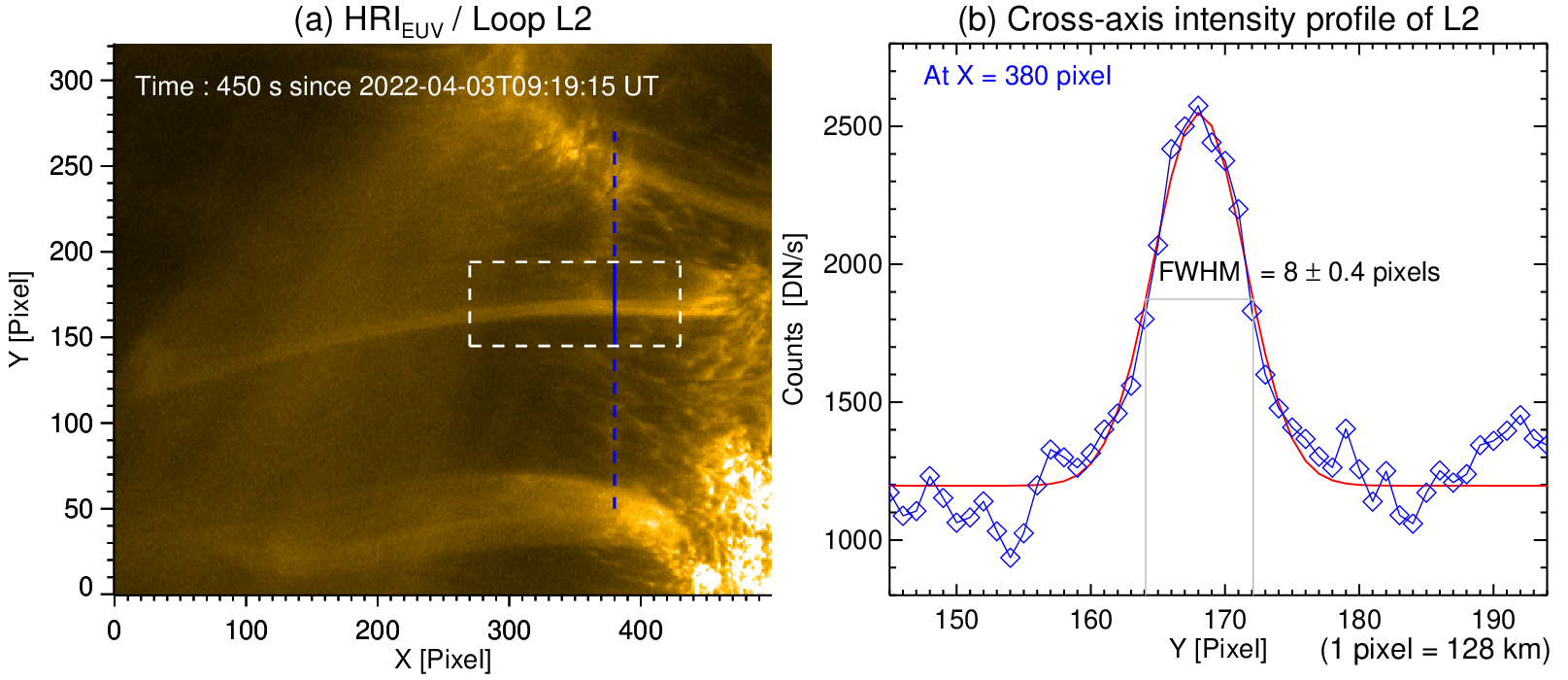}   
    \vspace{0.5cm} 
    \includegraphics[width=0.496\textwidth]{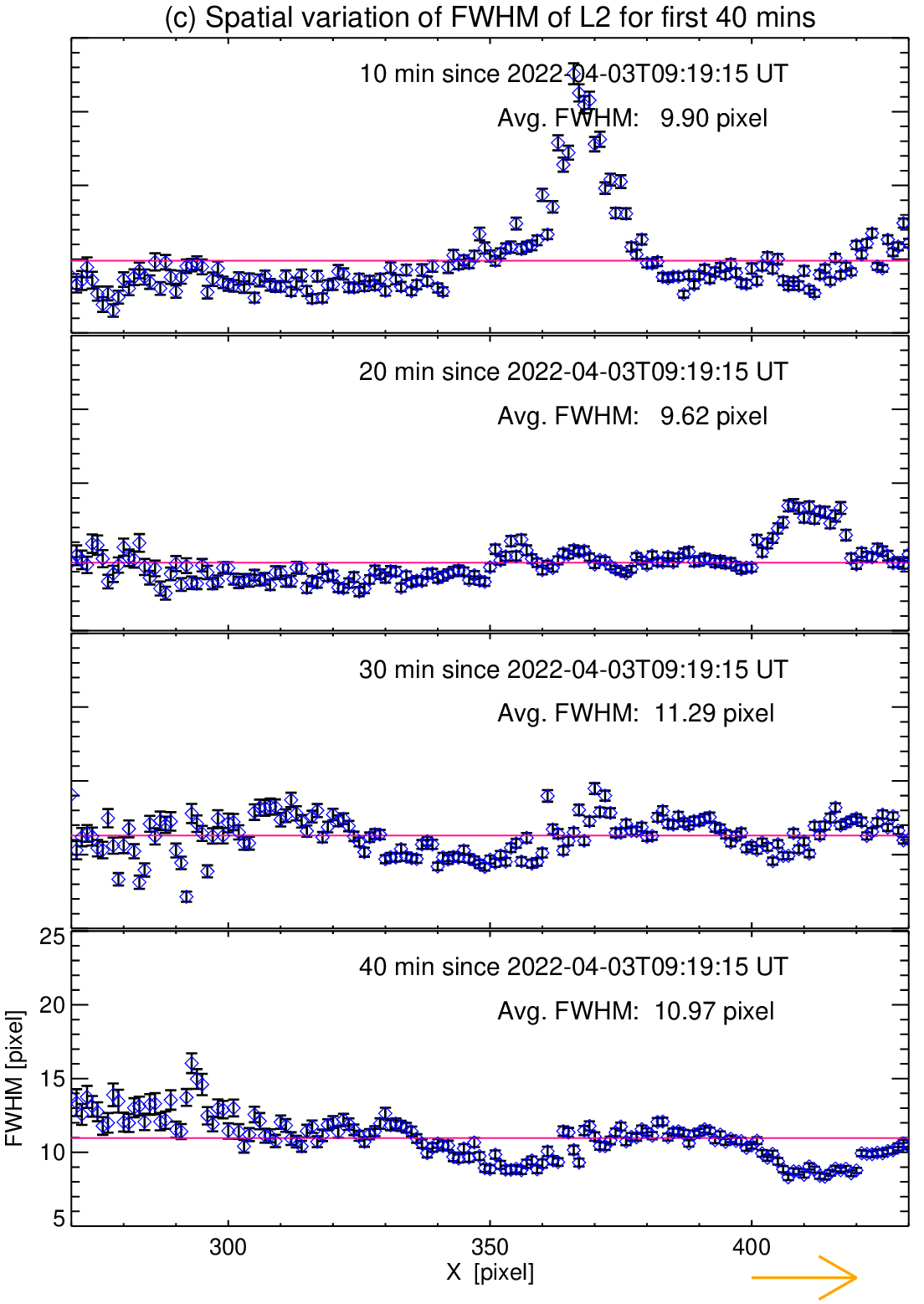} 
    \hfill
    \includegraphics[width=0.496\textwidth]{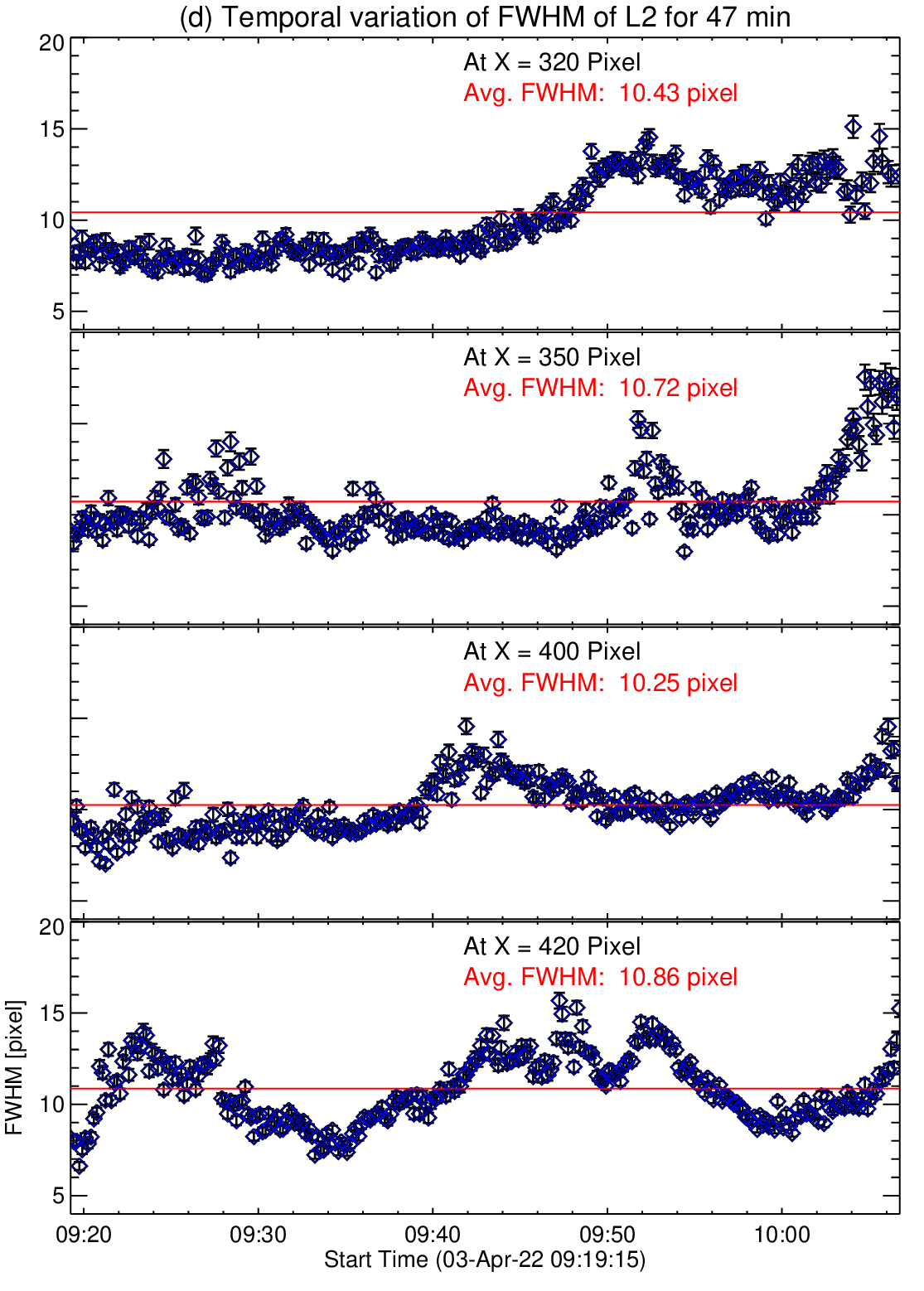}
    \caption{Same as Fig.~\ref{Fig2} but for loop L2}
    \label{Fig3}
    \end{figure*}

We followed the same method as used above now for loop L2. Most of the cross-section profiles of L2 also portray the smooth and single peaked nature similar to loop L1. An example of this is shown in Fig.~\ref{Fig3}(b). The cross-sectional profile is generated along the blue line drawn perpendicular to L2 at the position X=380 pixel (Fig.~\ref{Fig3}(a). The width (FWHM) of L2 at this position is measured as 8 pixels ($\approx$ 1 Mm). The HRI$_{\rm EUV}$ image of L2 is rotated 90\degr~anticlockwise of the region enclosed by the green square in Fig.~\ref{Fig1}(e). The size of the image is reduced in the vertical direction in order to have a close-up view of L2. 

The spatial variation of the width of L2 is investigated for the first 40 minutes along its length of about 20.5 Mm from X $=$ 270 to 430 pixels (equivalent to the dashed white rectangle length in Fig.~\ref{Fig3}a). The spatial variations of the FWHM of L2 measured at an interval of 10 minutes are plotted in four different panels of Fig.~\ref{Fig3}(c). In the top panel, the enhancement of the FWHM in L2 between X $=$ 360 and 375 pixels is attributed to transient, co-spatial background brightenings. These produce localized bulges, most evident around X $=$ 370 pixels (see attached movie), where overlapping emission broadens the CSI profiles and leads to larger measured widths. Such effects are not indicative of intrinsic loop expansion. Apart from this localized influence, the loop L2 width remains almost constant with an average of 10.4 $\pm$ 0.4 pixels ($\approx$ 1.3 Mm).

The lifetime of L2 is $>$ 50 minutes. However, L2 tends to diffuse after 47 minutes of observations. Hence, we investigated the temporal variations of L2 width for 47 minutes at every 10 pixels (1.28 Mm) distance along its length of 20.5 Mm. As sample plots, the temporal variations of L2 width at X $=$ 320, 350, 400, and 420 pixels are shown in Fig.~\ref{Fig3}(d). The apparent width enhancements of L2 at certain times, which produce the wavy pattern in Fig.~\ref{Fig3}d, are also linked to such transient background brightenings. They are associated with the periodic motions of EUV jet-like features, visible as parabolic tracks in the time–distance map (Fig.~\ref{FigA2}, Appendix~\ref{app2}). The overlapping emission from these features broadens the CSI profiles and mimics loop widening, but does not reflect genuine expansion of the loop. Overall, the width of L2 remains nearly constant along its length for most of its lifetime, suggesting a stable cylindrical tube-like morphology.

Interestingly, quite a number of cross-sectional profiles of L2 exhibited substructural complexities that provide additional information about its internal structure (see Fig.~\ref{Fig5} and Sect.~\ref{res2}). This does not contradict our conclusion about the nearly constant width of L2. The substructural complexities are confined to a short interval (09:26–09:33~UT), during which the cross-sectional profiles temporarily shift from single-peaked to double-peaked shapes. Importantly, the FWHM values remain essentially unchanged compared to the rest of the loop’s lifetime. Thus, the substructure reflects short-lived deviations in profile shape rather than significant changes in loop width.

\subsubsection{Loop width distributions}
\label {loop.width}
\begin{figure*}
        \centering
        \includegraphics[width=\hsize]{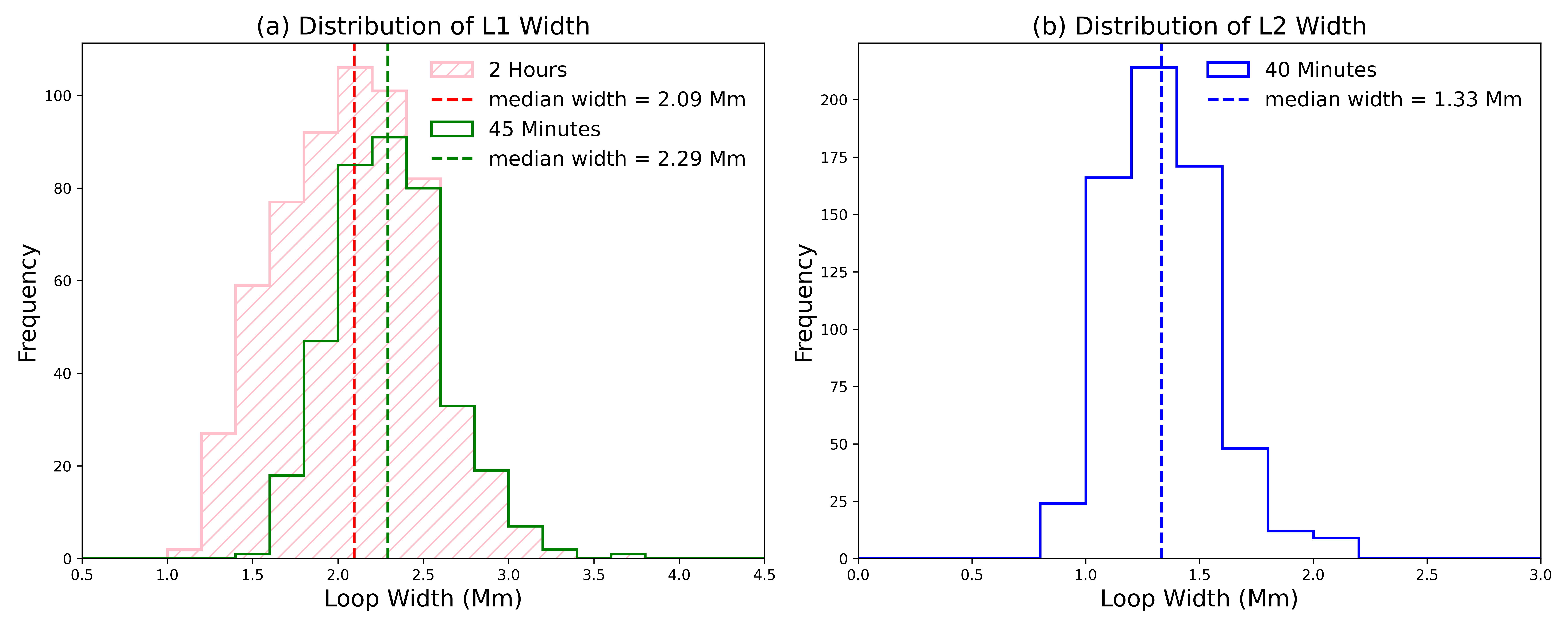}
        \caption{Histograms of width distributions of L1 (a) and L2 (b) with bin size of 0.2 Mm.}
        \label{Fig4}%
    \end{figure*}

The histograms of the width distributions of L1 and L2, derived from measurements along their respective lengths of 21.85 Mm and 20.5 Mm over observation periods of 120 minutes and 40 minutes, are shown in Fig.~\ref{Fig4}(a) and (b). To evaluate whether these distributions follow a normal distribution, we performed the Shapiro-Wilk test. The test results indicate that both distributions slightly deviate from normality (p < 0.05). Therefore, we used the inter-quartile range (IQR) statistics to compute the 1$\sigma$ uncertainty (IQR/1.35) and median as the measure of central tendency. Consequently, the computed 1$\sigma$ uncertainties for the median widths of L1 and L2 are 2.09 $\pm$ 0.45 Mm and 1.33 $\pm$ 0.21 Mm, respectively. It should be noted that the mean widths of L1 and L2 are very close to their medians, i.e., for L1, the mean is 2.08 Mm and for L2, the mean is 1.34 Mm. To assess how much the median (mean) of the L1 width distribution changes from the first 45 minutes to the full 2-hour period, we plotted the histogram for the first 45 minutes (in green) over the histogram for the entire 2-hour period (in pink), as shown in Fig.~\ref{Fig4}(a). The median (mean) width for the first 45 minutes is 2.29 (2.3) Mm, which is within the 1$\sigma$~uncertainty range of the 2-hour distribution. Thus, we statistically conclude that the L1 width remains nearly constant, with a median width of 2.1 $\pm$ 0.5 Mm over 2 hours. Similarly, the L2 width has a median of 1.3 $\pm$ 0.2 Mm for 40 minutes.

\subsection{Internal structure of the loops}
\label{res2}
\begin{figure*}
        \centering
        \includegraphics[width=\hsize]{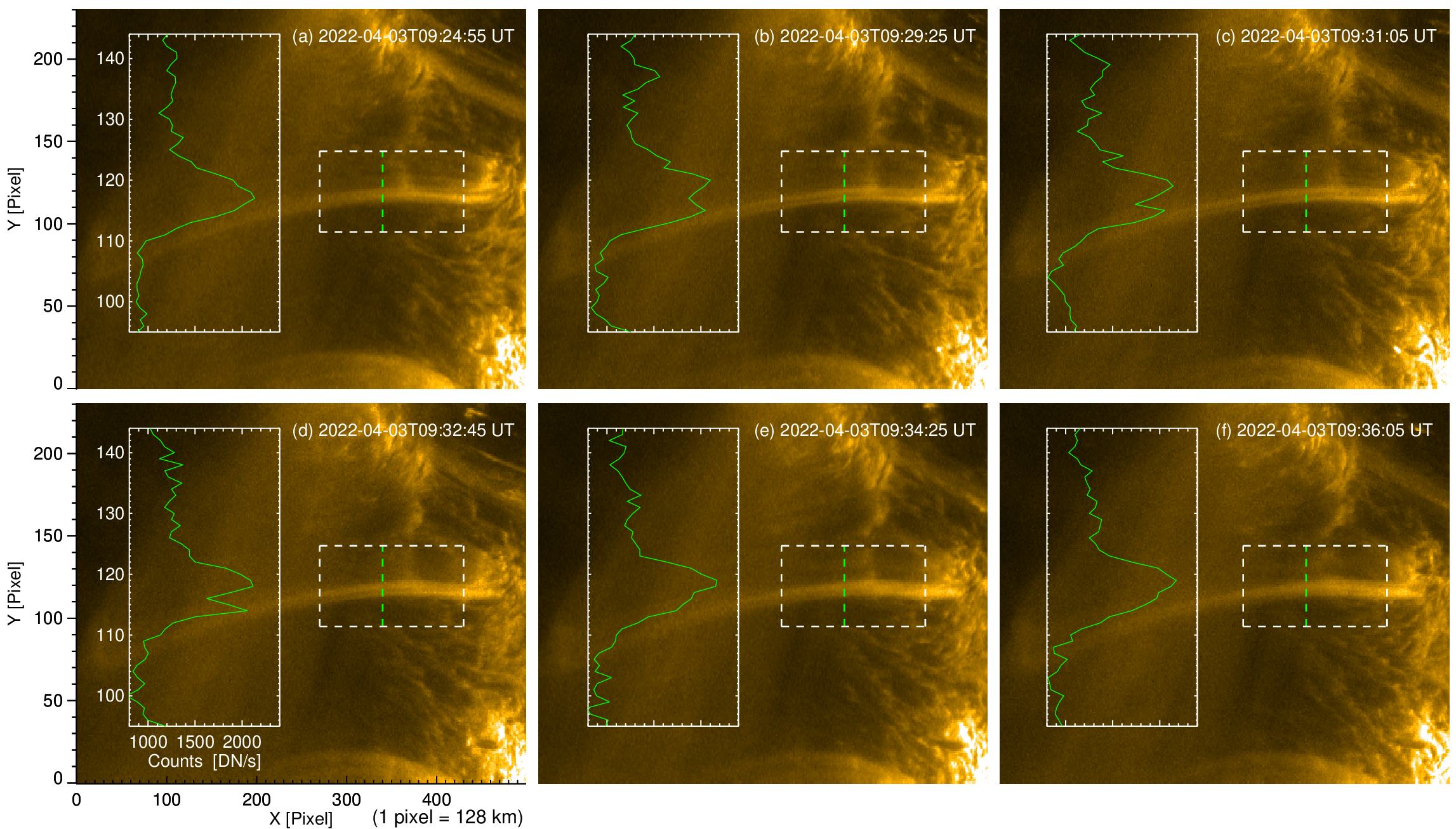}
        \caption{Observations indicating the absence of braiding in loop strands of L2. (a)  HRI$_{\rm EUV}$ image of L2 and the inset displays a uniform, single-peaked cross-sectional intensity profile (green solid curve) extracted along the green dashed line across L2. (b–d) A group of dim strands is observed running parallel to the adjacent bright strands on either side, with no evidence of entwined structures; note that the substructure of L2 is clearly resolved as a double-peaked profile in the cross-sectional intensity plots. (e–f) An apparent transition of the group of dim strands in L2 becoming bright, yielding a uniform single-peaked cross-sectional profile similar to that observed previously (a). }
        \label{Fig5}
    \end{figure*}

Throughout its evolution, loop L2 generally displays a single-Gaussian cross-sectional intensity profile, consistent with a relatively uniform and monolithic structure. However, during a short and rare interval between 09:26 UT and 09:33 UT on 3 April 2024, L2 exhibits a significant and transient departure from this pattern. As shown in Fig.~\ref{Fig5}(b–d), the cross-axis intensity profiles of L2 during this 7-minute window evolves into a distinct double-peak structure. A dimming of the emission is observed along the central axis of the loop, spanning nearly its full length. After 09:33 UT, this non-uniformity fades, and the loop gradually returns to a Gaussian-like profile (Fig.~\ref{Fig5}e-f), similar to its earlier state (Fig.~\ref{Fig5}a). 

This episode is notable because the dimming feature (defined here as a localized reduction in emission relative to the surrounding loop brightness, while remaining above the background level, and therefore not a completely non-radiating feature) persisted for about 7 minutes and remained spatially coherent over at least 25~Mm along the loop. During this interval, both the loop width and total intensity amplitude remained nearly unchanged (see Appendix~\ref{app5}). Moreover, the emission dip spanned about 3–4 pixels, corresponding to a spatial scale of approximately 384–512~km. The dip gradually deepened from 09:28:25~UT to 09:32:45~UT (Fig.~\ref{FigA5}b), indicating an evolving, internally heterogeneous structure rather than a homogeneous feature.

These observational characteristics argue against the possibility that the observed double-peak is caused by two independent, monolithic loops with a gap in between. If that were the case, we would expect changes in the overall loop width or intensity amplitude during their separation or coalescence. However, no such changes are seen. Instead, the structural coherence and profile stability suggest that L2 is a single-loop structure composed of two bright strand bundles with dimmer strands interspersed between them, giving rise to the observed double-peak intensity profiles. The temporal evolution of L2’s CSI profiles, transitioning smoothly from single-peak to double-peak and back to single-peak forms (Fig.~\ref{FigA5}, Appendix~\ref{app5}), together with a peak separation of 3-4 pixels that satisfies the Nyquist criterion, further supports internal structuring within a single loop. Nevertheless, we acknowledge that, given the finite spatial resolution, the alternative scenario of two nearly coaligned loops that momentarily diverge in brightness or position cannot be entirely excluded.

This group of dim loop strands extends nearly parallel to the adjacent bright loop strands in L2 for over 25~Mm, without any signs of localized brightenings or entanglements, which are typically associated with loop strand braiding \citep{Cirtain2013,Chitta2022}. The absence of braiding, at least down to the spatial resolution of HRI$_{\rm EUV}$ (256~km), within L2 raises several intriguing questions, which will be discussed in the next Sect.~\ref{dis}.

\section{Discussion}
\label{dis}
Coronal loops often exhibit Gaussian-like cross-sectional intensity profiles (e.g., \citealp{Peter2012}). However, the definition of loop width is not unique. For instance, \citet{Klimchuk2000a} adopted the second moment of the intensity profile as a characteristic measure of loop width. In this study, we used the FWHM of a Gaussian fit as a measure of loop width, as it is less sensitive to low-intensity wings and background variations. Loops L1 and L2 are longer than 40 Mm (twice the analyzed length of the loops), have diameters of about 2.09 Mm and 1.33 Mm, respectively, which is consistent with previous studies (eg. \citealp{Peter2013}). Although L1 and L2 are well resolved, with minimum widths consistently exceeding 6 pixels, their intensity profiles are mostly smooth, single peaked, and uniform. At few (short) time-intervals, L2 exhibits clear sub-structure profiles, allowing us to examine its internal structure effectively. This was mainly made possible by the high spatial resolution of HRI$_{\rm EUV}$.

The nearly uniform thickness of coronal loops along their lengths, indicative of missing or only weak expansion from footpoint to apex and a circular cross-section, has been reported in several previous studies \citep{Klimchuk2000a,Watko2000}. Recent studies, such as \citet{Sudip2024} and \citet{Ram2025}, utilizing stereoscopic observations, confirm that this non-expansive nature of loops is three-dimensional and not influenced by line-of-sight effects. All these studies focus on loops with lifetimes consistent with coronal cooling timescales of approximately 20-30 minutes (e.g., \citealp{Peter2012}). Quite recently, \citet{Li2020} analyzed a coronal loop with a lifetime of about 90 minutes and found similar evidence of only weak expansion along its length. However, an explicit study of width variations along the length of coronal loops throughout their extended lifetimes remains absent in previous studies. In this work, we investigated the width variations of two loops, L1 and L2, in detail along their lengths throughout their extended lifetimes, two hours and 50 minutes, respectively. We explicitly demonstrated that the loops maintain an almost constant width both along their lengths and over time. Our analysis raises several intriguing questions: What processes sustain these loops for such extended durations? What mechanisms promote their uniform thickness throughout their lifetimes? Could twisted magnetic flux tubes facilitate the observed uniformity? We explore possible explanations for these questions in the following subsections.

\subsection{Longevity of loops}
\label{dis1}

Coronal loops L1 and L2 exhibit distinct lifetimes and evolutionary patterns compared to their neighboring loops in the peripheral regions of ARs (see Appendix~\ref{app4}). The prolonged lifetimes of L1 and L2 should be governed by mechanisms different from those of nearby loops. In general, coronal loops are expected to cool rapidly due to efficient radiative and conductive losses, as indicated by energy loss calculations \citep{Cargill1994}. However, despite these theoretical expectations, observations show that coronal loops often persist longer than the typical cooling timescales (e.g., \citealp{Li2020}). To investigate this discrepancy, we computed the densities of loops L1 and L2 using both theoretical predictions based on the Rosner-Tucker-Vaiana (RTV) scaling laws \citep{Rosner1978} and direct observational measurements (see Appendix~\ref{app1}). Our analysis reveals that the observed densities are consistently higher than those predicted by RTV scaling laws \citep{Rosner1978}. This could indicate that L1 and L2 are not in static equilibrium, or they might be low-lying loops (far from semi-circular). Since the radiative cooling timescale is inversely proportional to density ($t_{rad} \propto 1/n_e$), the higher observed densities should, in principle, lead to even faster radiative cooling. However, the fact that these loops persist much longer than the cooling timescales suggests that additional heating mechanisms must be compensating for the rapid energy losses. And the very slow evolution of the loops indicates that this long-term energy input must be either constant or at a high frequency.

Several heating scenarios can account for the observed density enhancement and extended loop lifetimes. We investigated three potential heating mechanisms in our loops: 

\paragraph{(1) Footpoint-Driven Heating.} Loops experiencing strong heating near their footpoints (e.g., \citealp{Froment2017}) can develop evaporative plasma upflows from the chromosphere, leading to increased loop densities. In such a scenario, loops are expected to be in thermal non-equilibrium (TNE), characterized by a density decrease along the loop and periodic intensity pulsations indicative of recurrent evaporation and condensation cycles. However, the density distribution observed in L1 does not exhibit a gradient from its footpoint up to a distance of approximately 25 Mm (see Fig.~\ref{FigA1}(c)). Additionally, the symmetric brightness distribution along the loop and the heating dependency on electron density squared ($H \propto n_e^2$) suggest a more uniform rather than an asymmetric heating profile. Therefore, we consider footpoint-driven heating to be highly unlikely in loop L1. In contrast, in loop L2 a clear density gradient is observed (Fig.~\ref{FigA1}(d)). However, no evidence of intensity pulsations is found throughout its evolution (Fig.~\ref{FigA4}(d)). Consequently, we cannot confirm the presence of footpoint-driven heating in either loop. 

\paragraph{(2) Wave Heating.} One of the commonly observed wave signatures in coronal loops is that of slow magnetoacoustic waves. These waves manifest as propagating intensity disturbances in imaging data and are capable of transporting significant energy flux into the loop \citep{Nakariakov2000}. To identify such wave activity, we placed slits along loops L1 and L2 and constructed space-time plots (Fig.~\ref{FigA2}). However, no clear observational signatures of slow magnetoacoustic waves were detected in either loop (see Appendix~\ref{app2}). Another possible wave mode includes transverse waves, such as torsional Alfv{\'e}n waves and kink waves. Torsional Alfv{\'e}n waves are incompressible and can only be identified through their Doppler velocity signatures, which are not available in the present dataset. Kink waves, although also largely incompressible, cause periodic transverse displacements of the loop axis as they propagate \citep[e.g.,][]{Sudip2022}. To search for such motions, we generated X-T maps (not shown) by placing slits perpendicular to loops L1 and L2, away from their footpoints. However, no transverse motions were detected in either loop.

\paragraph{(3) Impulsive nanoflare heating.} According to \cite{parker1983,Parker1988} numerous small-scale impulsive events release energy locally somewhere within the loop envelope. The cluster of such small-scale reconnection events occurs stochastically in a very short time duration in a loop, leading to an intense energy release or "storms" of nanoflares \citep{Klimchuk2023}. The frequency with which these nanoflares occur in a loop determines its longevity. Specifically, if nanoflares occur more frequently than the loop's characteristic cooling timescale (20-30 min, \citealp{Peter2012}), the energy input can sustain coronal temperatures over extended periods. Conversely, if the frequency is lower than the cooling timescale, the loop cannot maintain its thermal structure and becomes transient. These episodic heating events can be detected as rapid and intense brightenings at the footpoints of hot loops \citep{Testa2013}. We investigated potential signatures of heating episodes due to nanoflares in loop L2 only, since the footpoint region of loop L1 appears saturated in the HRI$_{\rm EUV}$ images. This effect arises from the longer exposure time (2.8 s) used on 21 August 2021, compared to the typical 1.4–1.65 s, which caused saturation of pixels in the bright active-region core from which L1 fans out. The saturation is therefore instrumental in origin rather than a physical property of the loop. Appendix~\ref{app3} provides further methodological details and associated plots. We observed frequent intensity enhancements at the footpoint of loop L2, with fluctuations exceeding 20\% above the average normalized intensity, $\langle I/I_{\rm avg} \rangle$. As shown in Fig.~\ref{FigA3}(b), the time intervals between successive intensity enhancements are consistently shorter than 30 minutes. However, similar fluctuations are also present in nearby moss regions (Appendix~\ref{app3}, Fig.~\ref{FigA3}), indicating that these enhancements may be influenced by the background moss. Therefore, while high-frequency nanoflare heating remains a plausible mechanism for loop L2, these observations cannot unambiguously confirm its direct role in sustaining the loop. This is broadly consistent with the recent study of \citet{BMondal2025}, which showed that peripheral AR loops can be heated by high-frequency nanoflares.

Generally, nanoflares are thought to occur as a result of relaxation of braided loop strands in a loop. However, our observations suggest that the loop strands in loop L2 are unlikely to be braided with each other (Sect.~\ref{res2}). How nanoflares can occur in such a non-braided loop system as L2 will be discussed in the next Sect.~\ref{dis2}.

\subsection{``Non-braided'' multi-stranded loops}
\label{dis2}
Photospheric flows driven by turbulent convection influence the footpoints of coronal loop strands. Translational motions results in braiding, whereas rotational motions induce twisting of individual strands \citep{Klimchuk2015}. However, due to current instrumental limitations, directly resolving individual strands and their dynamics remains challenging. Simple numerical experiments by \citet{Peter2013} suggest that multi-stranded loops can produce uniform cross-sectional intensity profiles as they observed with the High-resolution Coronal Imager \cite[Hi-C;][]{2014SoPh..289.4393K}. They conclude that the width of individual strands in the smooth Hi-C loops (total loop width $\approx$ 1.5 Mm) cannot exceed 15 km. Using a similar approach, we estimate here that the individual strand diameters in our HRI$_{\rm EUV}$ observations are constrained to $\leq$ 35 km for loop L1 and $\leq$ 20 km for loop L2. The collective behavior of thousands of such fine strands could result in a smooth, uniform appearance over time.

Notably, in loop L1, some strands at the loop’s edge disappeared after 45 minutes of observation. In loop L2, between 09:28 UT and 09:33 UT, a group of dim strands emerged parallel to the bright strands without any observable signs of braiding at a resolution of  about 250 km. Since the estimated (maximum) strand size is approximately 20 km, which is much smaller than HRI$_{\rm EUV}$'s spatial resolution, we cannot determine whether braiding is absent at scales of a few tens of kilometers. However, our observations allow us to conclude that braiding is absent at spatial scales comparable to or larger than 250~km in the loops studied here. There are only a few reports on direct observations of braiding on such scales. Braiding was first reported by \citet{Cirtain2013} at such small scales using Hi-C data. Given the comparable spatial resolution of EUI, such braiding should be readily observed. However, \citet{Chitta2022} report that only a few cases of braiding are detected and conclude that braiding is a sparse phenomenon at the spatial scale resolvable by EUI ($\approx$ 250~km).

The observed absence of braiding at resolvable scales within L2 provides two key insights into its structure. First, L2 is not a monolithic entity but consists of numerous fine strands, providing observational evidence in support of multi-stranded coronal loop models (e.g., \citealp{Warren2002}). Second, it is highly unlikely that these loop strands are braided with each other or that the loop itself exhibits significant twisting along its length. This supports models (e.g., \citealp{Petrie2008}) suggesting that the magnetic twist of the flux tube enclosing the loop plasma is unlikely to be responsible for its nearly constant cross-section. The implications of this non-braided structure raise several intriguing questions. One such question concerns the apparent contradiction between the lack of braiding at observable scales and the presence of localized heating signatures.

For instance, the likely presence of nanoflare-like signatures at the footpoint of loop L2 (Appendix~\ref{app3}) suggests that individual strands may have twisted beyond the critical angle for kink instability, triggering impulsive reconnection. This interpretation aligns with 3D MHD models of multi-threaded coronal loops \citep{Cozzo2025}, which indicate that once a strand undergoes kink instability, it can merge with neighboring stable strands, producing an avalanche. During this phase, both ohmic and viscous dissipation contribute to the release of magnetic energy in discrete bursts, consistent with nanoflare theory.

There are two primary theoretical arguments linking braiding to loop width. The first argument suggests that if the braiding of thin strands in a loop corresponds to a twisted magnetic flux tube, the twist in the magnetic field lines introduces an azimuthal component. This component influences the overall magnetic pressure and tension balance, potentially counteracting the loop's natural tendency to expand because of decreasing external pressure with height \citep{Torok2003}. Under certain conditions, this mechanism can contribute to maintaining a nearly circular cross-sectional area along the length of the loop \cite{Klimchuk2000b}. The second argument posits that the braiding of thin strands induces magnetic reconnection between strands along the length of the loop. This process facilitates the cross-field dispersion of plasma across the loop, ultimately leading to the expansion of the coronal loop over its characteristic lifetime \citep{Schrijver2007}. However, our observations (Sect.~\ref{res2}) reveal no signatures of braiding at resolvable scales in any of the studied loops. As outlined above, it might be that very thin braided strands exists within the loop, but then the question would remain why there would be braiding only on scales below those observable without any noticeable signatures on the scales EUI can resolve. We therefore conclude that braiding and magnetic twist are unlikely to explain the constant cross-section observed in these loops.

\section{Summary and Conclusions}
\label{summ}
In this study, we analyzed the cross-sectional properties of two distinct coronal loops (L1 and L2) using high spatial resolution observations from HRI$_{\rm EUV}$ onboard the Solar Orbiter. These observations provide EUV images of L1 and L2 at spatial scales of 230 km per pixel and 128 km per pixel on the Sun, respectively. Both loops, located in the peripheral regions of active regions (ARs), are longer than 40 Mm and exhibit unusually long lifetimes, persisting for more than 2 hours (L1) and 50 minutes (L2), far exceeding the typical loop lifetime of 20 to 30 minutes.

We explicitly demonstrated the spatial and temporal variation of loop widths of L1 and L2 over their extended lifetimes. Our analysis reveals that both loops maintain a nearly constant cross-section throughout their evolution, indicating a non-expanding structure. The average widths of L1 and L2 are 2.1 $\pm$ 0.5 Mm and 1.3 $\pm$ 0.2 Mm, respectively, spanning about 9 to 10 pixels each. Furthermore, throughout their evolution, the loop widths consistently exceed 6 pixels, confirming that they are well resolved and exhibit uniform emission.

Based on a simple model, we estimate that individual strand diameters are constrained to $\leq$\,35~km in L1 and $\leq$\,20~km in L2. In L2, a brief episode reveals fine structuring, with faint strands running parallel to brighter ones, yet without any evidence of braiding or localized brightenings at the instrument’s spatial resolution of 256~km. These observations provide evidence that braiding or twisting is absent at spatial scales comparable to or larger than $\sim$250~km in the loops studied here.

Our results present contrarian evidence to the widely held idea that a twisted magnetic flux tube is required to maintain a uniform cross-section along a coronal loop. However, since our conclusions are based solely on emission observations, further confirmation requires high spatial, temporal, and spectral resolution measurements. Instruments such as the Multi-slit Solar Explorer (MUSE,~\citealp{Bart2022}) could provide Doppler shift measurements of loop strands, offering definitive evidence on whether braiding is present in coronal loops.

Additionally, we investigated potential heating mechanisms that could contribute to the extended lifetimes of these loops. In L2, rapid intensity fluctuations are detected near its footpoint, but their strong association with the background moss suggests that they cannot be unambiguously attributed to the loop itself. While high-frequency nanoflare heating remains a plausible mechanism, its direct role in sustaining L2 requires further confirmation. By comparison, L1 has a nearly uniform density, which indicates a more uniform heating process. However, it remains uncertain whether the temperature is also uniform, leaving open questions about the exact heating mechanism at play.

We conclude that the loops studied here are long non-expanding structures with uniform cross-section made up of numerous tiny strands. However, it remains unclear how these structures maintain their shape for such a long time in the corona, despite being part of an expanding magnetic environment. This leads to an important question: What physical processes contribute to the remarkable stability of these loops? While these long-lived smooth loops are rather rare specimen, their simple presence poses a problem not easy to answer by current models that usually concentrate on the highly-structured and dynamic nature of the solar corona.


\begin{acknowledgements}
We thank the referee for constructive comments that helped improve the quality of the paper. The work of N.V. and S.M. was supported by the Deutsches Zentrum f{\"u}r Luft und Raumfahrt (DLR; German Aerospace Center) by grant DLR-FKZ 50OU2201.
L.P.C. gratefully acknowledges funding by the European Union (ERC, ORIGIN, 101039844). 
\end{acknowledgements}

\bibliographystyle{aa}
\bibliography{ref.bib}

\begin{appendix}

\onecolumn
\section{Multi-strandedness nature of L2}
\label{app5}
\begin{figure}[htbp]
\centering
\includegraphics[width=\linewidth]{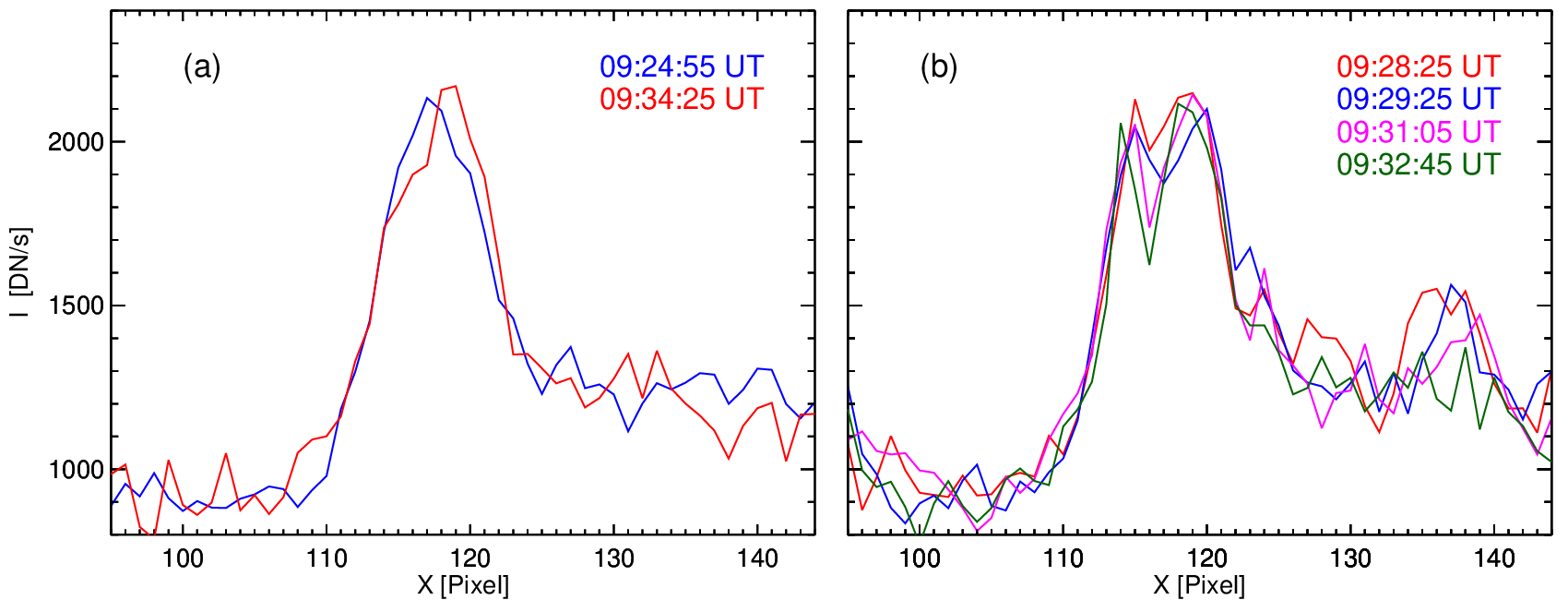}
\caption{(a) Cross-sectional intensity profile of L2 before the onset of the dimming (blue) and after the subsequent rebrightening (red). Note that both width and amplitude of the profile remain nearly constant. (b) Typical cross-sectional profiles overlaid with colors corresponding to their respective times, showing a gradual deepening of the intensity dip attributed to the evolution of dimming strands. The dimming strands spans (peak-to-peak distance) about 3-4 pixels (384-512 km).}
\label{FigA5}
\end{figure}

L2 exhibited asymmetric cross-sectional intensity profiles due to the presence of a dimming feature that persisted for approximately 7 minutes, from 09:26 UT to 09:33 UT on 3 April 2024. During this interval, both the width and amplitude of L2's cross-sectional intensity profile remained essentially constant. The intensity profiles prior to the onset of dimming and after the subsequent rebrightening are shown in Fig.~\ref{FigA5}(a). The representative intensity profiles over this time interval, overlaid in Fig.~\ref{FigA5}(b), further confirm the constancy of the loop’s width. The dimming feature began to rebrighten shortly after 09:33 UT.

This observation unambiguously rules out the possibility that the dimming feature was merely an intensity gap between two monolithic loops. In such a scenario, splitting or merging of loops would be expected to produce noticeable variations in the envelope width and/or intensity amplitudes. Therefore, L2 must consist of at least a three-stranded structure, with the dimming feature located between two sets of bright strands, resolved at the spatial scale of 250 km. Moreover, the gradual deepening of the intensity dip from 09:28:25 UT to 09:32:45 UT (Fig.~\ref{FigA5}b) indicates that the dimming structure is not homogeneous, but instead comprises multiple strands that progressively transition into a more strongly dimmed state. The dimming feature spans approximately 3–4 pixels (corresponding to 384–512 km). Based on this, we infer that L2 likely possesses a multi-stranded configuration, with the dimming feature consisting of a collection of 19–25 dimming strands aligned along the line of sight, considering an individual strand diameter of $\leq 20$ km (see Sect.~\ref{dis2}).
\FloatBarrier

\FloatBarrier

\section{Light curves of loops}
\label{app4}
\begin{figure*}[h!]
    \centering
    \includegraphics[width=0.48\textwidth]{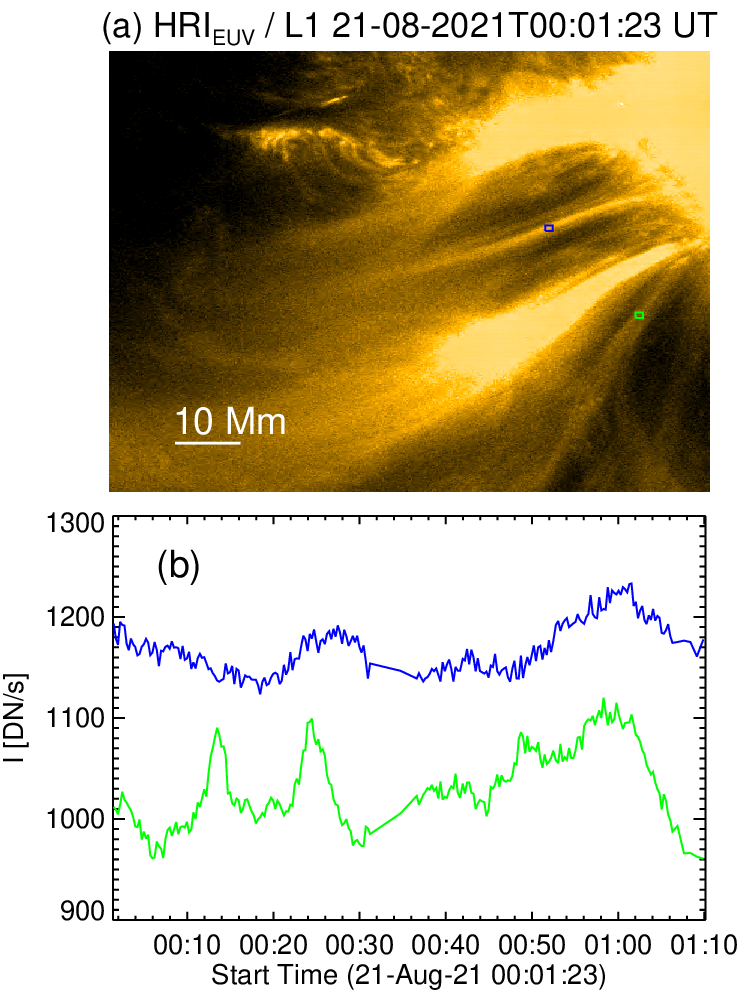}
    \includegraphics[width=0.48\textwidth]{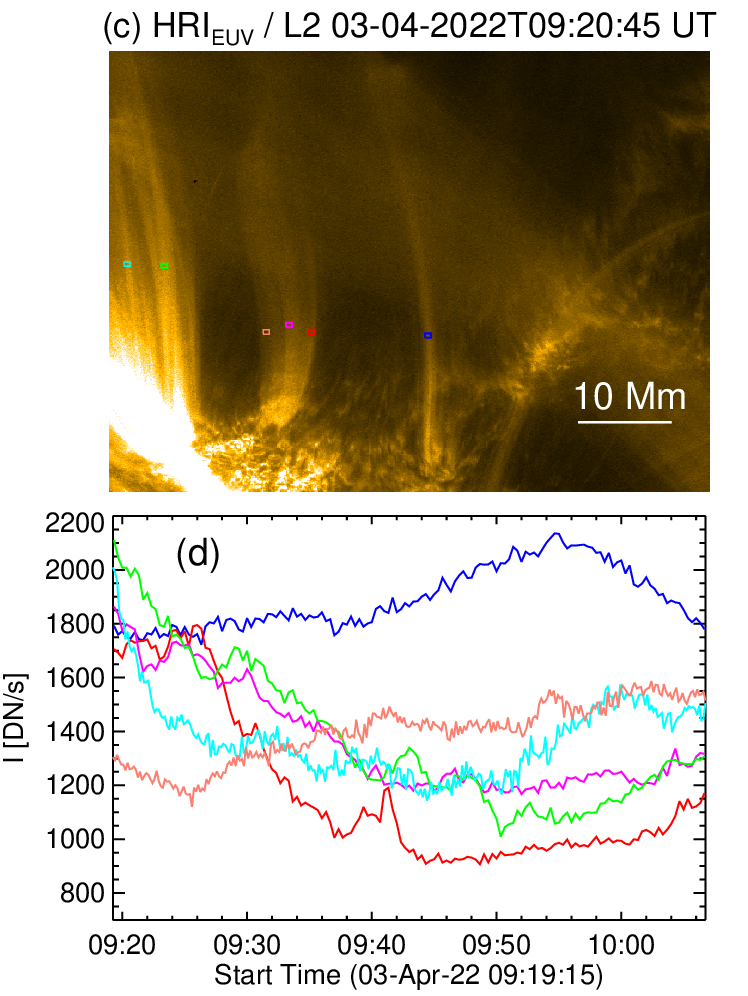}
    \caption{(a) A close-up view of the HRI$_{\rm EUV}$ region within the green box in Fig.~\ref{Fig1}(b), enclosing loop L1. The small boxes, each of size 5$\times$5 pixels marked in different colors, are used to compute the light curves, which are plotted in the corresponding colors in (b). (c)-(d) Same as (a)-(b), but for loop L2. Notably, both L1 and L2 (blue curves) exhibit minimal variations in intensity during their evolution unlike their neighboring counterparts.}
    \label{FigA4}
\end{figure*}

The evolution of loops L1 and L2 was studied in comparison with their neighboring loops using emission light curves, as shown in Fig.~\ref{FigA4}. The light curves of the loops were computed using the average intensity values over small regions of 5 $\times$ 5 pixels, marked with different color boxes as shown in Fig.~\ref{FigA4}(a) \& (c). As seen in Fig.~\ref{FigA4}(b), the emission from L1 (blue) exhibits no significant variations, whereas the emission from the neighboring loop (green) shows notable fluctuations. A similar behavior is observed for L2 (blue), where its emission remains stable compared to its neighboring loops, as shown in Fig.~\ref{FigA4}(d). The contrasting emission patterns of L1 and L2 relative to their neighboring loops suggest that the underlying physical mechanisms governing their evolution differ from those of the typical short-lived loops.

\FloatBarrier

\section{Density of loops}
\label{app1}
\subsection{From RTV scaling laws} 
Under static equilibrium and uniform heating along the loop, the relation between loop length (L), temperature (T) and pressure (p) is given by \citep{Rosner1978} ,
\begin{equation}
     T = 1.4 \times 10^3 (pL)^{1/3} ,
\label{eq1}
\end{equation}
where $T$ is in K, $L$ is in cm, and $p$ is in dyne/cm$^2$. 
Using the ideal gas law,
\begin{equation}
    p = 2 k_B T n_e,
\label{eq2}
\end{equation}
where $k_B = 1.38 \times 10^{-16}$ erg/K is the Boltzmann constant and $n_e$ is electron density.

On substituting equation~\ref{eq2} in~\ref{eq1} and solving for $n_e$,

\begin{equation}
    n_e = \frac{T^2}{2 k_B (1.4 \times 10^3)^3 L}
\label{eq3}
\end{equation}

At $T = 1$ MK,
\begin{equation}
    n_e = \frac{1.32 \times 10^{18}}{L}
\label{eq4}
\end{equation}

As the exact length of our loops is not known, densities of coronal loops are computed for the lengths ranging from 20 Mm to 120 Mm.

For $L = 20$ Mm in equation~\ref{eq4}, $n_e = 6.6 \times 10^8 \text{ cm}^{-3}$ 

\& for $L = 120$ Mm in equation~\ref{eq4}, $n_e = 1.1 \times 10^8 \text{ cm}^{-3}$

\subsection{From Observations} 

Since a single waveband does not provide full temperature distribution information, we made reasonable assumptions on Emission Measure (EM) to compute the density ($n_e$) of the coronal loops from HRI$_{\rm EUV}$. As we know the response function $R(T_{peak})$ of the HRI$_{\rm EUV}$ at $T_{peak} = 1MK$ is $10^{-23.6} DN cm^{-5} s^{-1} pxl^{-1}$ (Fig.1 in \citealp{Chen2021}). The observed intensity (I) can be defined as,
\begin{equation}
I = EM \cdot R(T_{peak})
\label{eq5}
\end{equation}
where EM is in cm$^{-5}$, I is in $DN s^{-1} pxl^{-1}$ and $R(T_{peak})$ is in $DN cm^{-5} s^{-1} pxl^{-1}$.

We assumed that the emitting plasma is isothermal at this peak temperature of 1MK, so the EM can be approximated to,
\begin{equation}
 EM = \int n_e^2 \, dl \approx  n_e^2 l 
 \label{eq6}
\end{equation}
where the line-of-sight depth $l$ is taken as the width of the loop measured.

On combining above two equations (\ref{eq5} \&~\ref{eq6}), the electron density $n_e$ can be computed as,
\begin{equation}
n_e = \sqrt{\frac{EM}{l}} =  \sqrt{\frac{I}{R(T_{peak}) \cdot l}}
\end{equation}

\begin{figure*}[h!]
    \centering
    \includegraphics[width=0.496\textwidth]{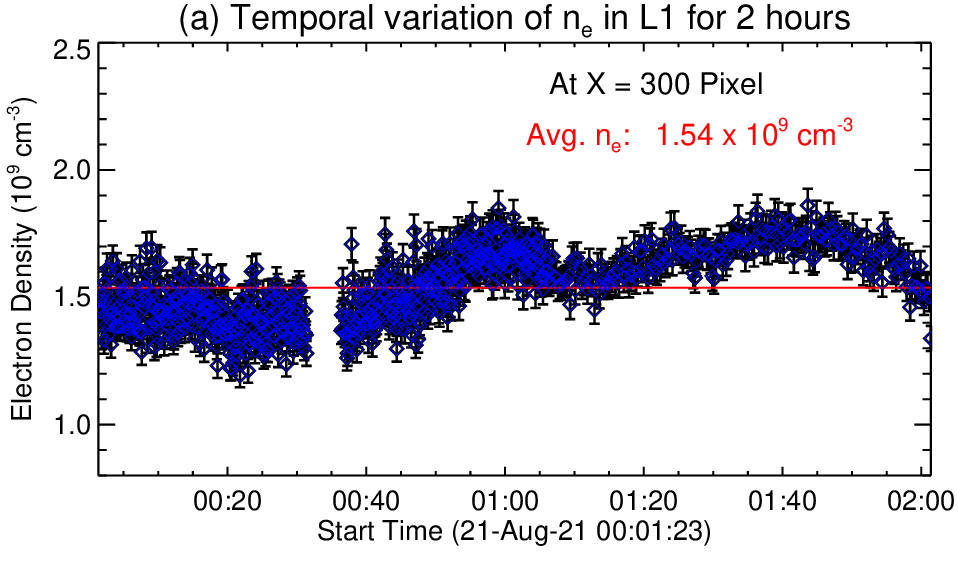}
    \hfill
    \includegraphics[width=0.496\textwidth]{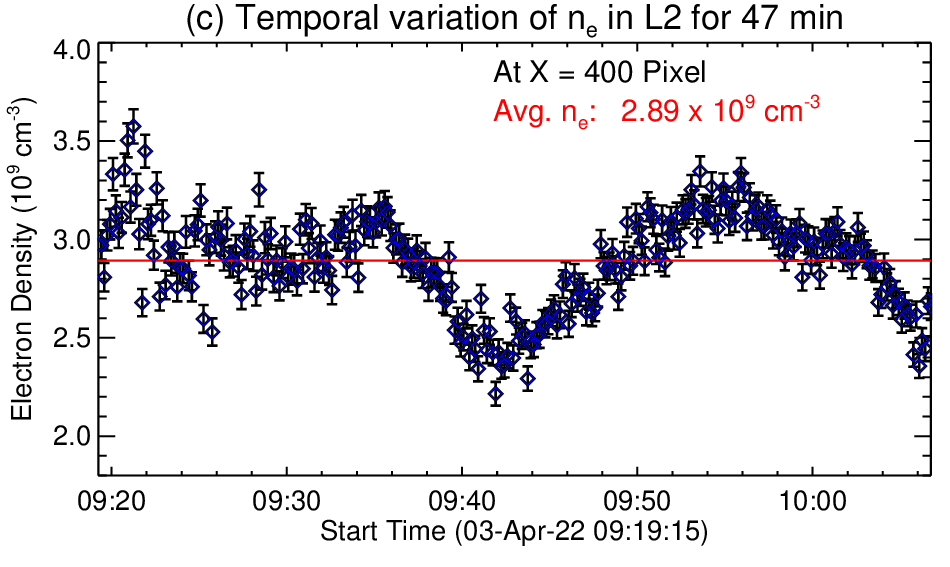}
    \vspace{0.5cm} 
    \includegraphics[width=0.496\textwidth]{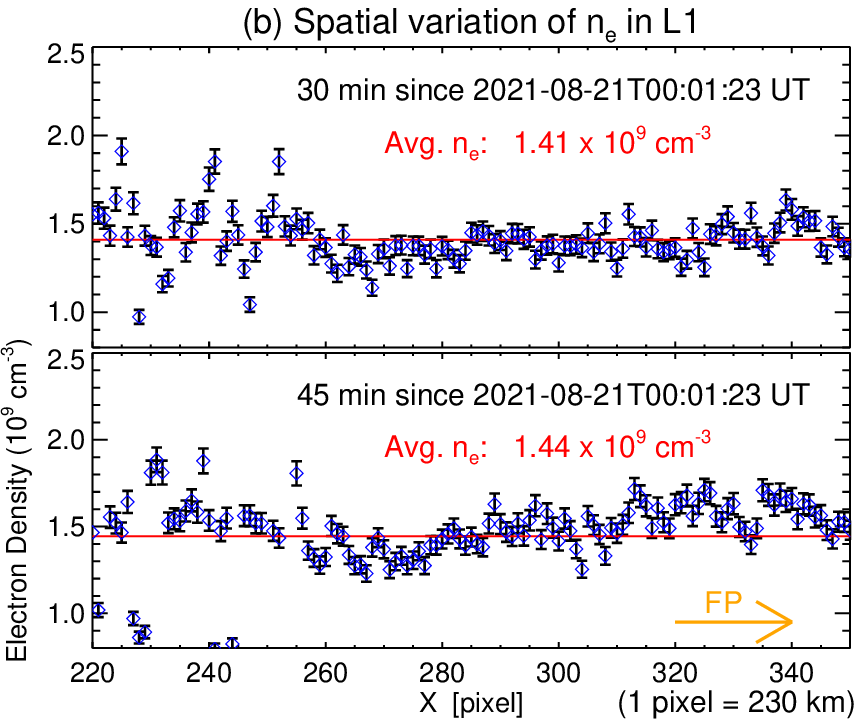}
    \hfill
    \includegraphics[width=0.496\textwidth]{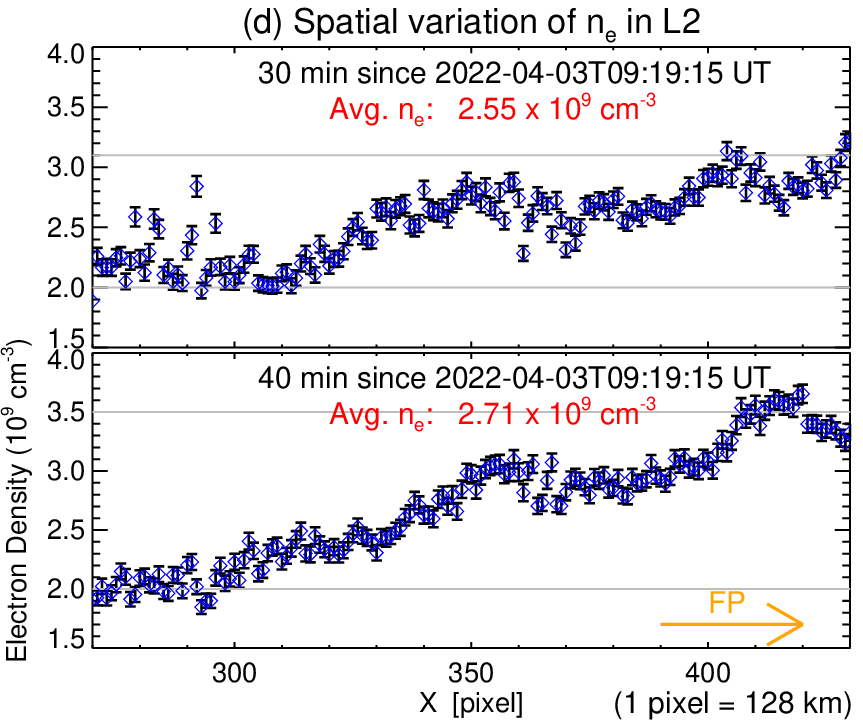} 
    \caption{ (a) The variations of $n_e$ in loop L1 at position X=300 pixel for over the time period of 2 hours. Horizontal line in red marks the mean $n_e$ over this time period. (b) Same as (a) but for loop L2 for over the time period of 47 minutes. (c) Uniform variation of $n_e$ along L1 after 30 and 45 minutes from the start time of L1 observations. (d) Spatial variation of $n_e$ along L2 after 30 and 40 minutes from the start time of L2 observations. Two Grey horizontal lines are plotted to highlight the increase in $n_e$ from higher part of L2 towards its footpoint (FP). Orange arrows point to loop footpoints.}
    \label{FigA1}
    \end{figure*}

The typical $n_e$ evolution for loops L1 and L2 over their lifetimes computed at the position X=300 and X=400 pixel is shown in Fig.~\ref{FigA1}(a) \& (c), respectively. The average densities of L1 and L2 at these positions are measured as $1.54 \times 10^9 cm^{-3}$ and $2.89 \times 10^9 cm^{-3}$, respectively. The observed densities are clearly over-dense compared to those predicted by RTV scaling laws, suggesting that the loops are continuously heated by mechanisms that balance conductive and radiative losses. Even when assuming a line of sight depth equal to half the loop width, which is appropriate given the cylindrical geometry of the loops, the estimated densities increase further, and it does not impact the results significantly for this study. Furthermore, investigation of spatial variation of $n_e$ along L1 and L2 reveals an interesting contrast (see Fig.~\ref{FigA1}(b)~\& (d)). In L2, $n_e$ increases toward the footpoint, indicating a clear density gradient along the loop. In contrast, L1 exhibits a remarkably uniform $n_e$ along its length. This suggests that L1 is heated uniformly along the loop, while L2, despite showing a density gradient, does not exhibit intensity pulsations during its evolution. Therefore, we cannot confirm whether L2 is in thermal nonequilibrium and heated by footpoint driven mechanisms.

\FloatBarrier

\section{Space-time plots}
\label{app2}
\begin{figure*}
        \centering
        \includegraphics[width=0.97\textwidth]{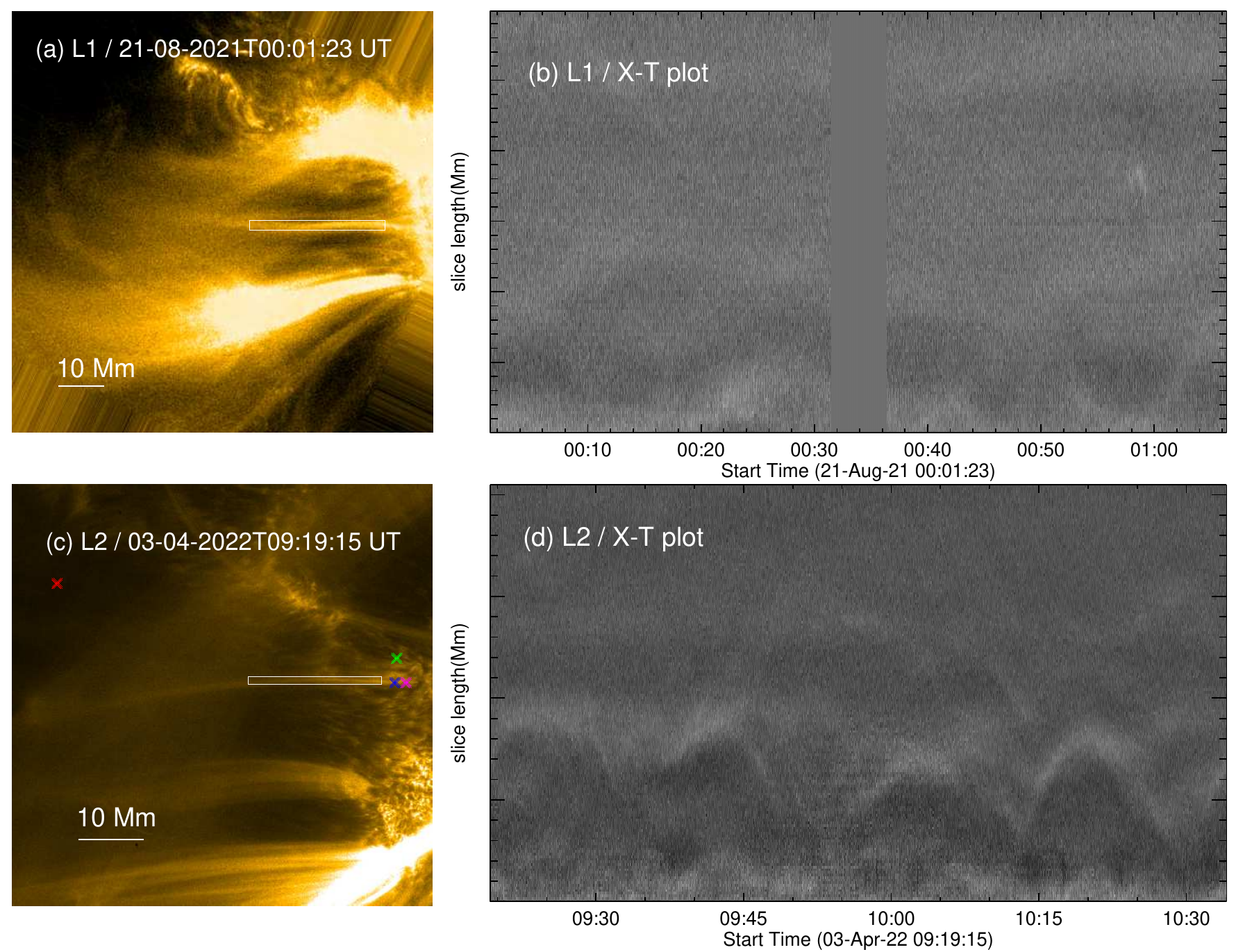}
        \caption{(a) Slit is placed along loop L1 and the corresponding space-time plot is displayed in (b). (c) \& (d) is same as (a) \& (b), but for loop L2. The cross symbols in (c) marks the location of four small sub-regions of 3 $\times$ 3 pixels, over which the normalized intensity variations are displayed in Fig.~\ref{FigA3}.}
        \label{FigA2}%
    \end{figure*}

To investigate the possible outward propagation of intensity fluctuations along the loops, we placed a slit along the loops, as shown in Fig.~\ref{FigA2} (a) \& (c). The slit widths are comparable to the loop widths. The space-time (X-T) plots generated from these slits are presented in Fig.~\ref{FigA2} (b) \& (d), do not exhibit any such propagation. The parabolic pattern observed in Fig.~\ref{FigA2}(d) is due to the periodic up-and-down movement of EUV bright jet-like features, which are the coronal counterparts of cooler chromospheric dynamic fibrils \citep{Sudip2023,2023A&A...678L...5M}. These jet-like features have a size of 3-4 Mm and oscillation period of about 10-15 min are part of the background variations and not associated with the loops. These X-T plots show no evidence of outward propagating intensity disturbances along the loops, which would otherwise support the continuous MHD wave heating scenario.

\FloatBarrier

\section{Footpoint variability}
\label{app3}

\begin{figure*}
        \centering
        \includegraphics[width=0.98\textwidth]{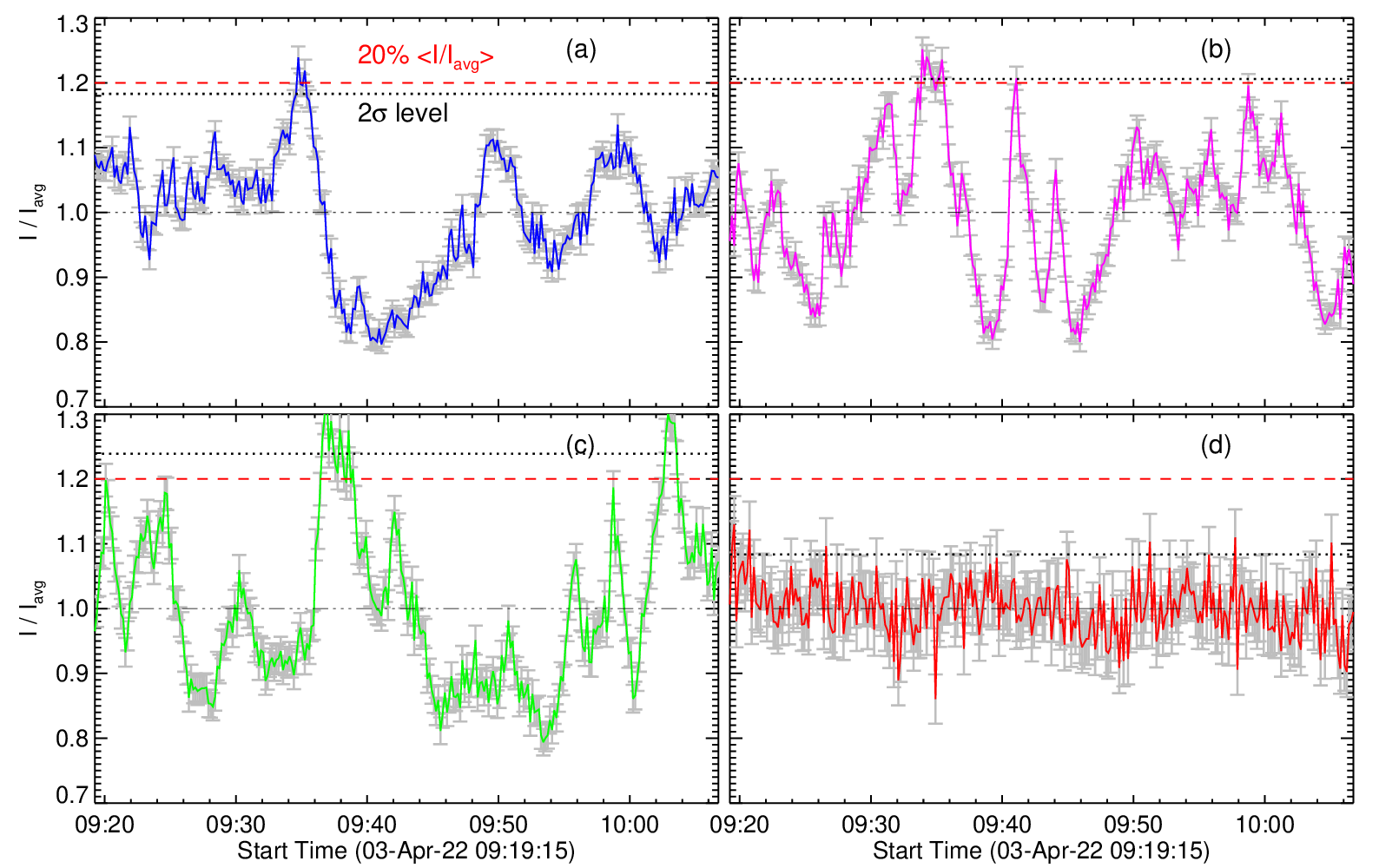} 
        \caption{Normalized light curves for the four regions marked in Fig.~\ref{FigA2}(c). (a) \& (b) Normalized intensity variations at two different locations within the EUV L2 footpoint. (c) Same as (a), but for a region in the moss. (d) Same as (c), but for a region farther from the moss. The threshold of 20$\% ~\langle I/I_{\rm avg} \rangle$ and 2$\sigma$ levels are indicated by dashed (in red) and dotted horizontal lines, respectively, in all panels.}
        \label{FigA3}%
    \end{figure*}

We investigated rapid intensity enhancements at or near the L2 footpoint to search for potential signatures of nanoflares using data with a 10-second cadence. The L1 footpoint falls within the saturated region of the AR and therefore, our analysis is limited only to L2. We followed the methodology of \citet{Testa2013} and selected four distinct regions, each measuring 3×3 pixels—comparable to the effective spatial resolution of HRI$_{\rm EUV}$, with each pixel corresponding to 128 km on the Sun. Two of these regions are located over the EUV footpoint of L2, the third is in the nearby moss region, and the fourth is farther away from the moss region. These regions are marked with four cross symbols of different colors in Fig.~\ref{FigA2}(c). The light curves were generated by integrating the intensity within these regions, then normalized to their respective averages, and plotted in Fig.~\ref{FigA3} using the same colors as their regions are marked. The threshold of 20$\% ~\rm I/I_{\rm avg}$ and 2$\sigma$ levels is overlaid in each panel to filter out spurious intensity enhancements. Intensity enhancements that reached the threshold of 20$\%$ of the average normalized intensity ($\langle I/I_{\rm avg} \rangle$) are considered potential signatures of heating episodes due to nanoflares in loops. In Fig.~\ref{FigA3}(a), we found one prominent intensity enhancement that reached the threshold level. However, as we go a little down towards the L2 footpoint, the frequency of intensity enhancements reaching the threshold level increases (Fig.~\ref{FigA3}b). Nanoflare-like intensity fluctuations are found in the moss-dominated footpoint segment of L2 (Fig.~\ref{FigA3}a, b), as well as in a nearby moss region away from the loop footpoint (Fig.~\ref{FigA3}c). In contrast, in a region located away from both the moss and the L2 footpoint (Fig.~\ref{FigA3}d), no comparable enhancements above the detection threshold are observed. This spatial dependence suggests that the observed fluctuations are most likely associated with the background moss, and therefore cannot be unambiguously attributed to loop L2 itself. Thus, while high-frequency nanoflare heating remains a plausible mechanism, its direct association with loop L2 cannot be firmly established.

\FloatBarrier

\end{appendix}
\end{document}